\documentclass[]{pasj01}
\draft

\Received{}
\Accepted{}

\usepackage[T1]{fontenc} 
\usepackage{lineno}

\begin{document} 

\title{
The correlation between the 500 pc scale molecular gas masses and AGN powers for massive elliptical galaxies}

\author{Yutaka \textsc{Fujita}\altaffilmark{1}%
}
\altaffiltext{1}{Department of Physics, Graduate School of Science
Tokyo Metropolitan University
1-1 Minami-Osawa, Hachioji-shi, Tokyo 192-0397}
\email{y-fujita@tmu.ac.jp}

\author{Takuma \textsc{Izumi}\altaffilmark{2}}
\altaffiltext{2}{National Astronomical Observatory of Japan, 2-21-1 Osawa, Mitaka, Tokyo 181-8588}

\author{Nozomu \textsc{Kawakatu}\altaffilmark{3}}
\altaffiltext{3}{National Institute of Technology, Kure College, 2-2-11, Agaminami, Kure, Hiroshima, 737-8506, Japan}

\author{Hiroshi \textsc{Nagai}\altaffilmark{2,4}}
\altaffiltext{4}{The Graduate University for Advanced Studies, SOKENDAI, Osawa 2-21-1, Mitaka, Tokyo 181-8588}

\author{Ryo \textsc{Hirasawa},\altaffilmark{1}}
\author{Yu \textsc{Ikeda}\altaffilmark{1}}


\maketitle

\begin{abstract}
Massive molecular clouds have been discovered in massive elliptical
galaxies at the center of galaxy clusters. Some of this cold gas is
expected to flow in the central supermassive black holes and activate
galactic nucleus (AGN) feedback. In this study, we analyze archival ALMA
data of 9 massive elliptical galaxies, focusing on CO line emissions, to
explore the circumnuclear gas. We show that the mass of the molecular
gas within a fixed radius (500~pc) from the AGNs ($M_{\rm mol}\sim
10^7$--$10^8\rm\: M_\odot$) is correlated with the jet power estimated
from X-ray cavities ($P_{\rm cav}\sim 10^{42}$--$10^{45}\rm\: erg\:
s^{-1}$). The mass accretion rate of the
circumnuclear gas $\dot{M}$ also has a correlation with $P_{\rm
cav}$. On the other hand, the continuum luminosities at $\sim 1.4$~GHz
and $\sim 100$--300~GHz have no correlation with $M_{\rm mol}$. These
results indicate that the circumnuclear gas is sustaining the long-term
AGN activities ($\sim 10^7$~yr) rather than the current ones. 
The circumnuclear gas mass is a better indicator of the jet power than the continuum luminosity, which probably changes on a shorter time scale. We also
study the origin of the continuum emission from the AGNs at $\sim
100$--300~GHz and find that it is mostly synchrotron radiation. For
low-luminosity AGNs, however, dust emission appears to contaminate the
continuum.
\end{abstract}

\section{Introduction}

Massive elliptical galaxies are often located at the center of galaxy
clusters. They host supermassive black holes, which are observed as
active galactic nuclei (AGNs). While the cooling time of the hot gas in
the galaxies is much smaller ($\lesssim 10^8$~yr) than the age of the
Universe \citep{1994ARA&A..32..277F}, the AGNs are expected to serve as
heating sources and prevent the development of massive cooling flows
(AGN feedback;
\cite{2000A&A...356..788C,2007ARA&A..45..117M,2012ARA&A..50..455F}).
Since the galaxies are immersed in the hot gas, it may be natural to
assume that the hot gas feeds into the black holes in the form of the
Bondi accretion \citep{1952MNRAS.112..195B} and stimulates the AGNs.  In
fact, \citet{2006MNRAS.372...21A} found a tight correlation 
between the Bondi accretion rates and the power emerging from the AGNs
in relativistic jets. However, later studies have revealed that the
correlation is weak \citep{2013MNRAS.432..530R}.

In addition to the hot gas, cold molecular gas has been discovered in
the massive elliptical galaxies
(e.g. \cite{2001MNRAS.328..762E,2003A&A...412..657S,2014ApJ...792...94D,2014ApJ...785...44M}). It
has been indicated that nebular emission, enhanced star formation, and
AGN activities are recognized when the entropy of the hot gas decreases
to $\lesssim 30\:\rm keV\: cm^2$, or almost equivalently when the
radiative cooling time is $\lesssim 1$~Gyr
(e.g. \cite{2008ApJ...683L.107C,2008ApJ...687..899R,2009MNRAS.398.1698S,2017MNRAS.464.4360M}). This suggests that radiative cooling turns the hot gas into cold gas, and the cold gas drives star formation and AGN activity. Recently, \citet{2022ApJ...924...24F} discussed gas
circulation in the massive elliptical galaxies using a semi-analytical
model. They indicated that while some of the cold gas is consumed in
star formation, the rest flows into the galactic center. Moreover, if a
circumnuclear disk is formed around the central black hole, disk
instabilities regulate the gas accretion toward the black hole; the
overall AGN activity depends on the mass of the disk. We note that the
typical size of the circumnuclear disk is $\sim 100$--500~pc for the
massive elliptical galaxies at the cluster centers
\citep{2019ApJ...883..193N,2022ApJ...924...24F}.

In this paper, we analyze Atacama Large Millimeter Array (ALMA) archival
data of massive elliptical galaxies in galaxy clusters and seek
correlations among the properties of circumnuclear molecular gas and the
AGN activity, which may support the idea that the gas fuels the AGNs. We
note that \citet{2019MNRAS.490.3025R} have compared the mass of the
circumnuclear molecular gas with the AGN jet power and found a
correlation (their Figure~7). However, they measured the mass in a
single ALMA synthesized beam centered on the AGNs. Since the physical
size of a given beam is larger for more distant objects, the mass inside
it tends to be larger, while distant AGNs are observationally biased
toward brighter and powerful ones. This could create an artificial
correlation. To lessen such biases, we focus on the mass within a fixed
physical radius. We also study the mass accretion rates and the radio
continuum luminosities of the AGNs to further investigate the fueling
process.  We assume $H_0=70\rm\: km\: s^{-1}\: Mpc^{-1}$,
$\Omega_m=0.3$, and $\Omega_\Lambda=0.7$ in this study. All errors are
$1\:\sigma$ unless otherwise noted.

\section{Data Reduction}

We analyzed archival ALMA observations of CO line emissions from the
cold gas in massive elliptical galaxies at the center of galaxy
clusters, and we investigated relations between the cold gas and the
AGN activity. Targets were selected from those studied in
\citet{2019A&A...631A..22O} and \citet{2019MNRAS.490.3025R}. From the
sample, we chose the ones for which AGN jet powers ($P_{\rm cav}$)
have been estimated through observations of X-ray cavities. Moreover, we
chose those for which the inner 500~pc (1~kpc in diameter) is resolved
with ALMA. The selected sample includes M87 and RXJ~0821$+$0752. However,
the molecular gas detected in M87 is located in the outer region of the
galaxy and has no connection to the AGN \citep{2018MNRAS.475.3004S}, and
the data of RXJ~0821+0752 are poorly calibrated because of an issue of
water vapor radiometer \citep{2017ApJ...848..101V}. Thus, we did not
include them in our sample. The list of the 9 remaining galaxies are
shown in table~\ref{tab:obs}. We call the central galaxies by the names
of their host clusters. The galaxy redshifts $z$ are those determined by
optical observations.

The galaxies were observed with ALMA at frequencies corresponding to the
CO(J=1-0), CO(J=2-1), or CO(J=3-2) rotational transition lines as well as at
additional spectral windows that were used to image the mm/submm ($\sim
100$--300~GHz) continuum emission.  The observations were single
pointing ones centered on the AGNs, and they were calibrated using the
appropriate version of Common Astronomy Software Application (CASA)
software \citep{2007ASPC..376..127M} and ALMA Pipeline used for the
quality assurance.  Data sets taken in the early ALMA Cycles were
calibrated by the EA ALMA Regional Center (EA-ARC) through the request
via the ALMA Helpdesk because of unavailability of the recommended CASA
version. 

The line and underlying continuum emissions were reconstructed using the
CASA \texttt{tclean} task down to $3\:\sigma$ level. To study the line
emissions, we subtracted the continuum using line-free channels with the
CASA task \texttt{uvcontsub} (fitorder=1).  Following previous studies
\citep{2019A&A...631A..22O,2019MNRAS.490.3025R}, we adopted Briggs
weighting with a robust parameter of 2 (natural weighting). Each
receiver was tuned to cover one of the abovementioned lines. The targets
were observed with four spectral windows except for 2A0335+096 and
PKS~0745-191 (two windows). Each spectral window has a bandwidth of
1.875 or 2.0~GHz. The synthesized beam size, and rms in each
final data-cube are shown in table~\ref{tab:obs}.

\begin{table}
  \tbl{Target and observation details.}{%
  \begin{tabular}{cccccccccc}
      \hline
Target &  $z$\footnotemark[$*$] & CO line & ALMA ID project & Obs. time & Date & Beam & PA & $v$ binning & rms\footnotemark[$\dag$] \\
 &  &  &  & (min) &  & ($''$) & (deg) & ($\rm km\: s^{-1}$) & ($\rm mJy\: bm^{-1}$) \\
      \hline
NGC 5044 & 0.00928$^a$ & J=2-1 & 2011.0.00735.S & 28 & 2012-01-13 & 2.1 $\times$ 1.2 & -33 & 20 & 1.4 \\
Centaurus & 0.0099$^b$ & J=1-0 & 2015.1.01198.S & 78 & 2016-01-27 & 2.3 $\times$ 1.8 & 76 & 30 & 0.3 \\
Abell  262 & 0.01619$^c$ & J=2-1 & 2015.1.00598.S & 11 & 2016-06-27 & 0.95 $\times$ 0.62 & 12 & 20 & 0.6 \\
Abell 3581 & 0.0218$^c$ & J=2-1 & 2015.1.00644.S & 87 & 2016-05-03 & 0.69 $\times$ 0.59 & -77 & 20 & 0.3 \\
Abell  2052 & 0.03455$^d$ & J=2-1 & 2015.1.00627.S & 80 & 2016-08-11 & 0.33 $\times$ 0.27 & -41 & 20 & 0.3 \\
2A0335+096 & 0.03458$^e$ & J=1-0 & 2012.1.00837.S & 71 & 2014-07-22 & 1.3 $\times$ 1.1 & -26 & 20 & 0.5 \\
Hydra A & 0.05435$^f$ & J=2-1 & 2016.1.01214.S & 45 & 2016-10-23 & 0.24 $\times$ 0.17 & 89 & 10 & 0.4 \\
Abell  1795 & 0.06331$^d$ & J=2-1 & 2015.1.00623.S & 70 & 2016-06-11 & 0.81 $\times$ 0.61 & -15 & 20 & 0.4 \\
PKS 0745-191 & 0.1028$^c$ & J=3-2 & 2012.1.00837.S & 25 & 2014-08-19 &  0.27 $\times$ 0.19 & 78 & 20 & 0.8 \\
      \hline
    \end{tabular}}\label{tab:obs}
\begin{tabnote}
\footnotemark[$*$] References of redshifts: ($a$) \citet{2008AJ....135.2424O}, ($b$) \citet{1992CORV..C...0000F}, ($c$) \citet{2019A&A...631A..22O}, ($d$) \citet{2017ApJS..233...25A}, ($e$) \citet{2019ApJ...872..134Z}, and ($f$) \citet{2019MNRAS.485..229R}.\\
\footnotemark[$\dag$]  A noise level for a channel map with a velocity width of $v$ binning.
\end{tabnote}
\end{table}

\section{Results}

Using the CASA task \texttt{immoments}, images of the integrated
intensity, intensity-weighted mean radial velocity (velocity
center hereafter), and intensity-weighted mean velocity dispersion (full
width at half-maximum or FWHM) were created for the vicinities of the
AGNs in the 9 galaxies covering each CO line
(figure~\ref{fig:image}). We extracted spectra from the central 500~pc
region centered on the AGNs. We confirmed that the maximum
recoverable scales are much larger than 500~pc for all the targets and
the missed flux from the 500~pc region should be ignorable. Each spectrum was
fitted with a model consisting of one or two Gaussian components
(figure~\ref{fig:spec}). The significance levels of those line
components are $>5\:\sigma$. The best-fitting line center velocity $v$
relative to the galaxy rest frame, FWHM, and integrated intensity for
each gaussian component $S_{\rm CO}\Delta v$ are presented in
table~\ref{tab:result}.  We ignored line-like features if the
line center velocity is $|v|\gtrsim 500\rm\: km\: s^{-1}$ because the
corresponding gases are unlikely to be bound to the nuclei although they
may still reside in the disks of central galaxies.  They may be gas
blobs that are falling toward or being ejected from the galactic
centers. Hydra~A has a prominent cold molecular gas disk seen
in CO and substantial absorption due to the disk is seen at the line
frequency at the position of the AGN (\cite{2019MNRAS.485..229R}; see
figure~\ref{fig:image}). Thus, we excluded the innermost 160~pc, and we
extracted and examined the spectra in the eastern and western regions of
the AGN separately (figure~\ref{fig:spec}). The parenthesized FWHM
values shown in table~\ref{tab:result} are the flux averages of the two
gaussian components except for Hydra~A, for which it is the flux average
of the east and west values.

\begin{figure}
 \begin{center}
  \includegraphics[width=17cm]{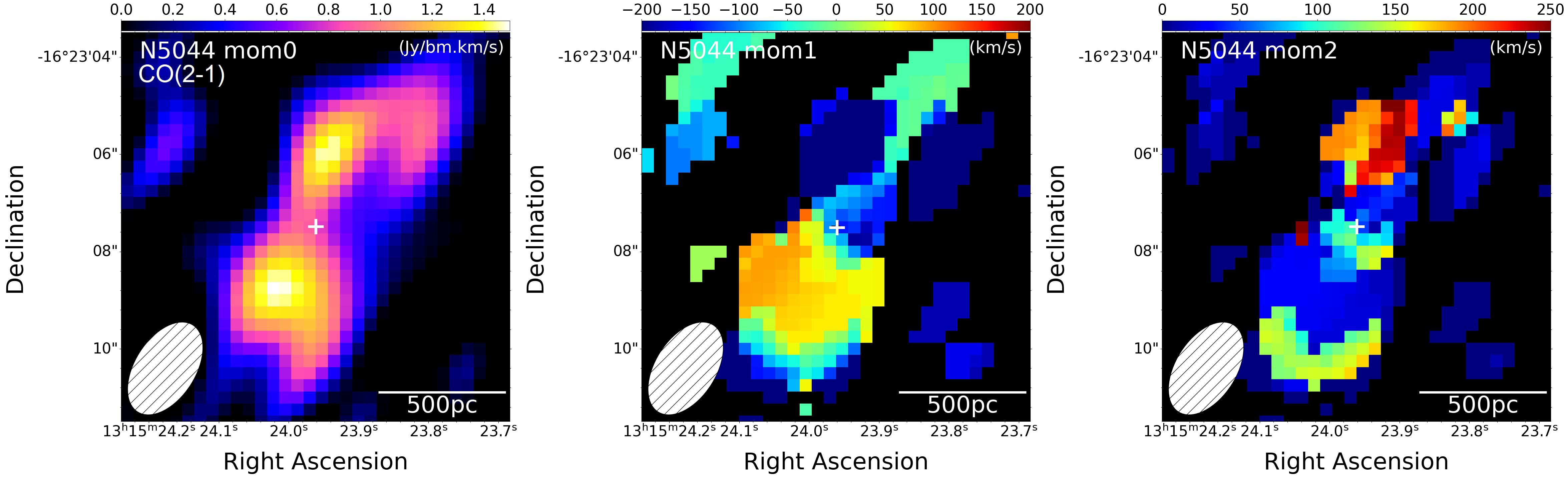} 
  \includegraphics[width=17cm]{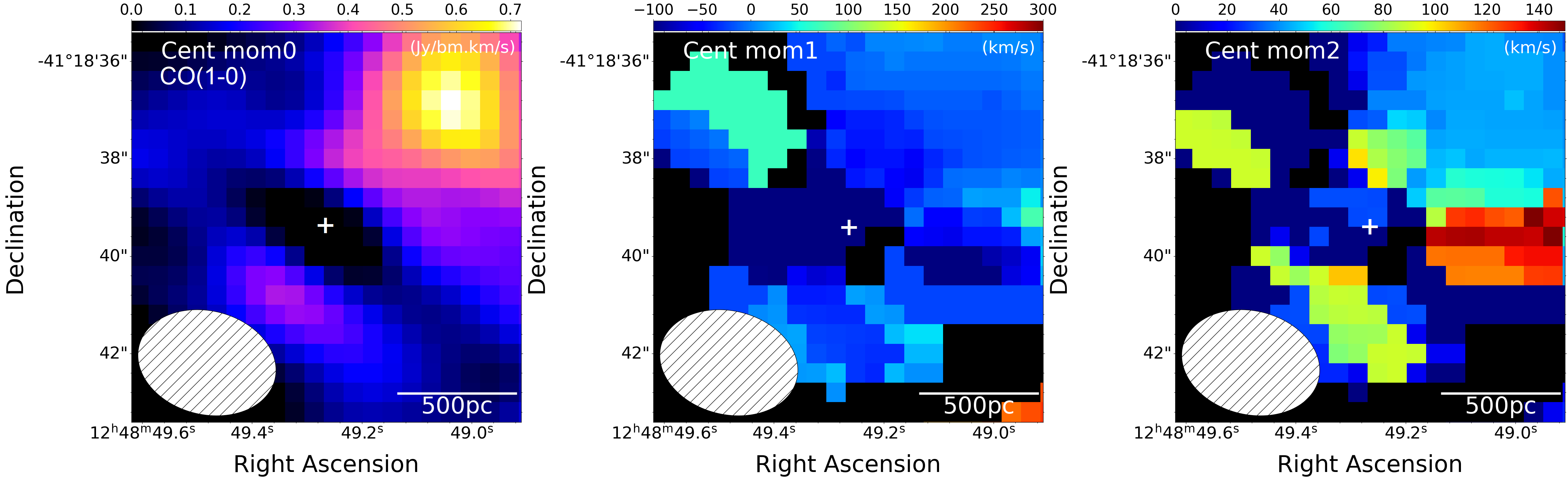}
  \includegraphics[width=17cm]{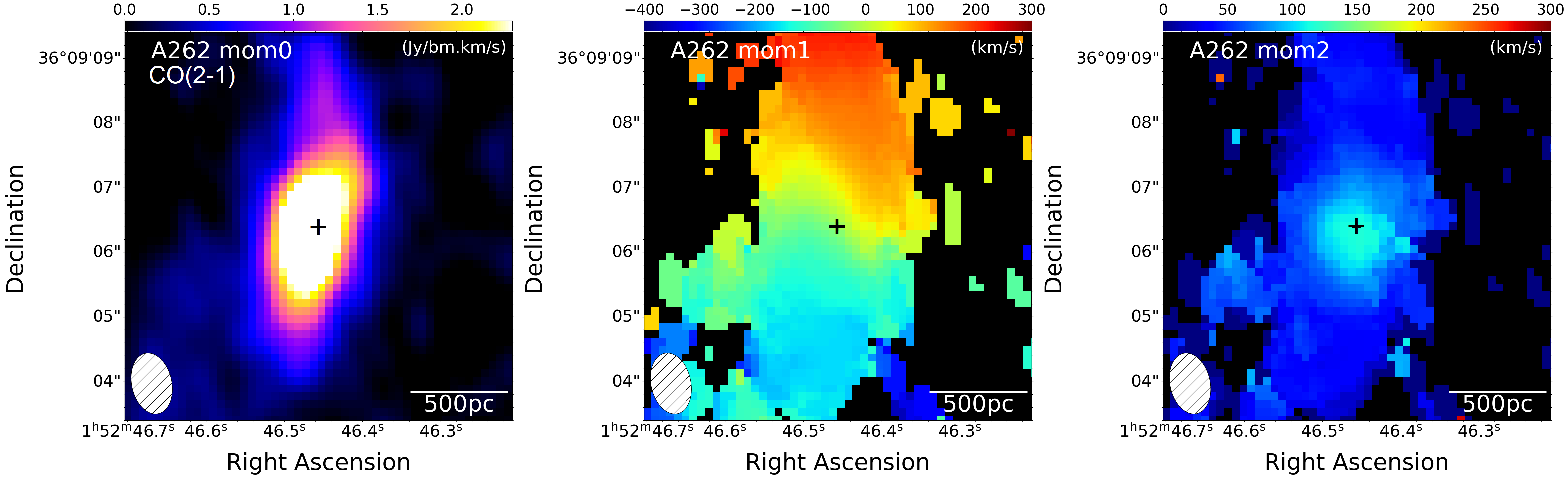}
  \includegraphics[width=17cm]{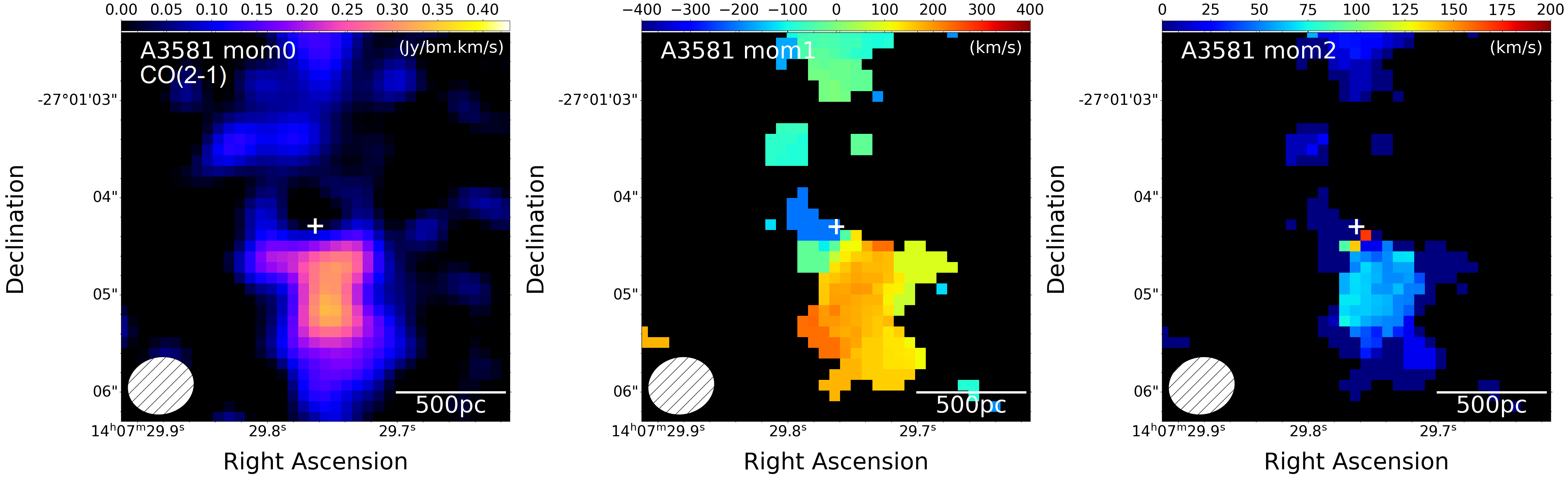}
 \end{center}
\caption{From left to right, close-up CO integrated intensity (moment 0)
map, velocity center distribution (moment 1) map of the $\geq 3\sigma$ CO
emission, and velocity dispersion (moment 2) map of the $\geq 3\sigma$
CO emission. The AGNs are located at the center of the figures. The beam
is plotted in white at the bottom left side of each panel. Note that the
maps for Abell~2052 appear to be different from those provided by
\citet{2019MNRAS.490.3025R}. This is may be because of the difference of
beam sizes.}\label{fig:image}
\end{figure}

\begin{figure}
 \begin{center}
 \addtocounter{figure}{-1}
   \includegraphics[width=17cm]{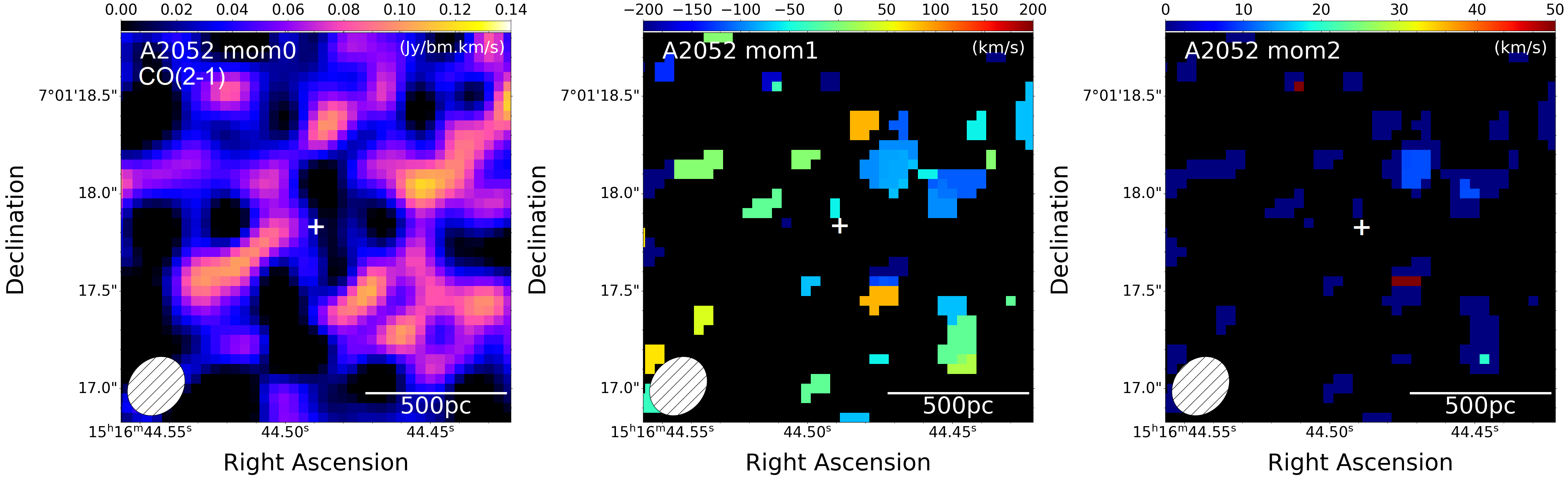}
   \includegraphics[width=17cm]{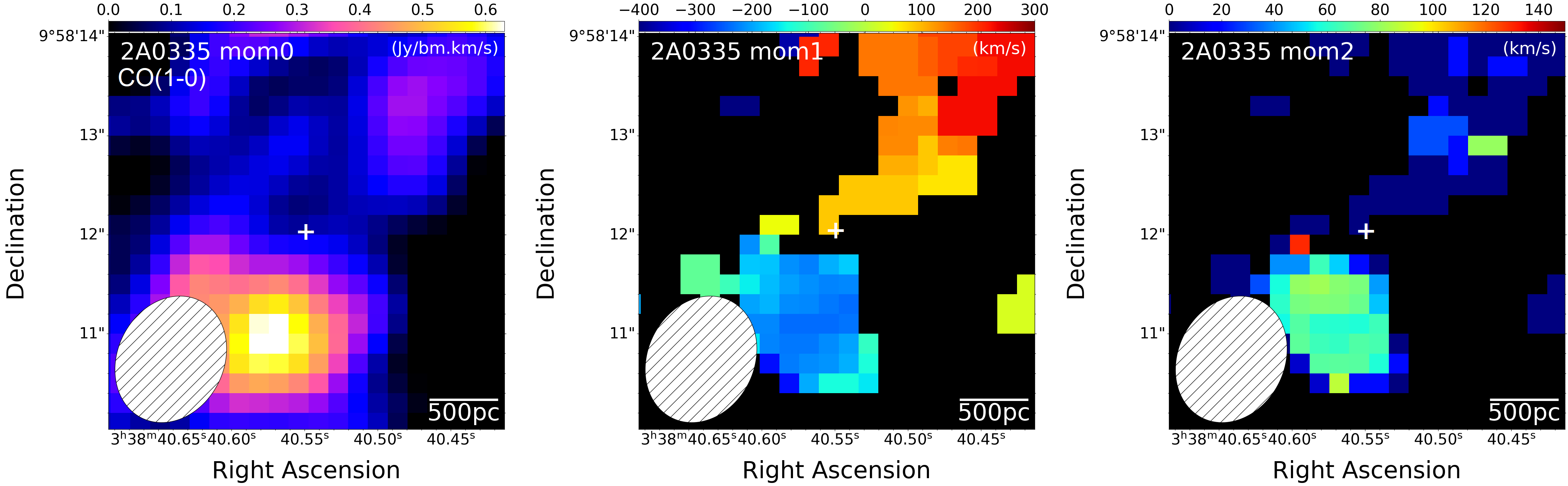}
   \includegraphics[width=17cm]{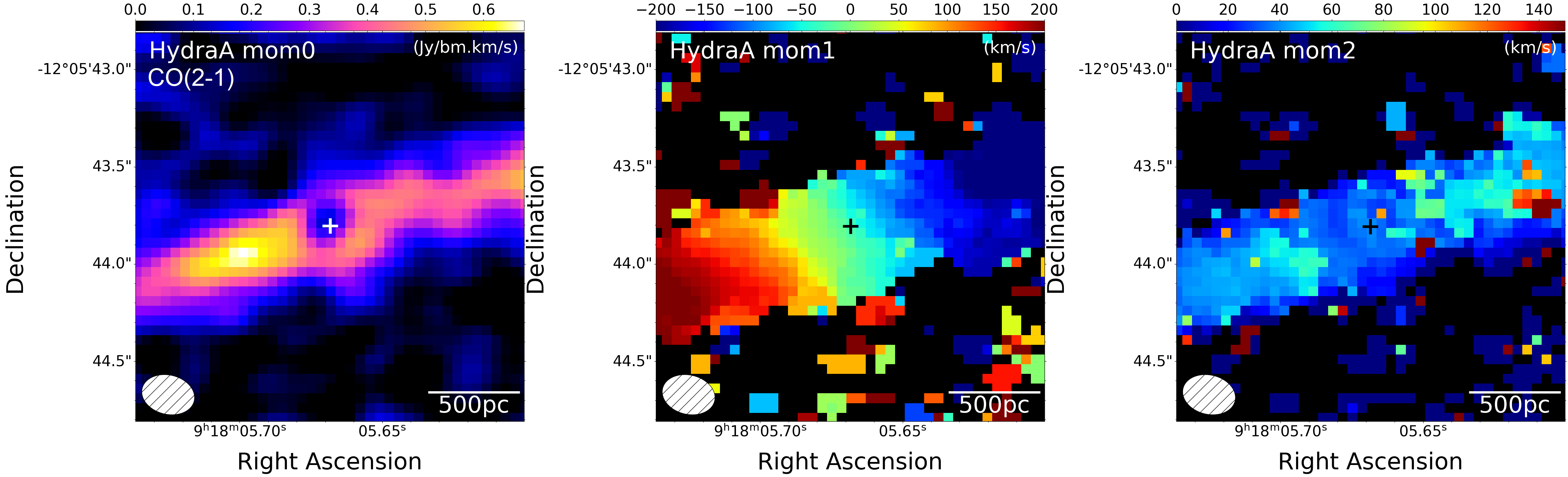}
   \includegraphics[width=17cm]{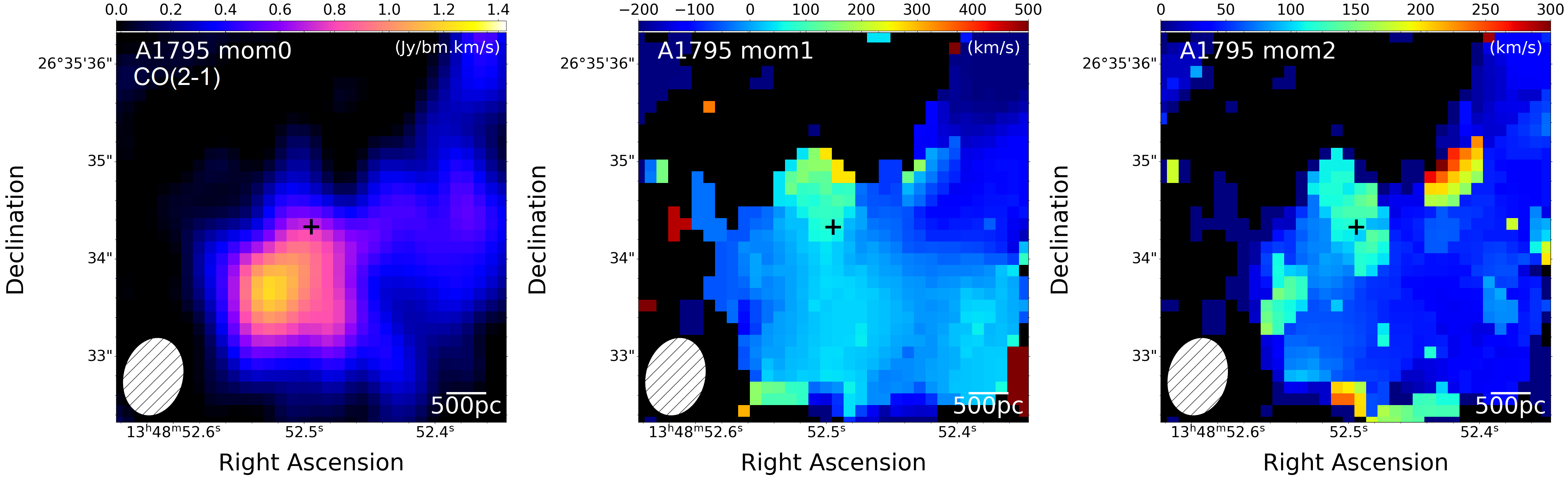}
 \end{center}
\caption{Continued}
\end{figure}

\begin{figure}
 \begin{center}
 \addtocounter{figure}{-1}
   \includegraphics[width=17cm]{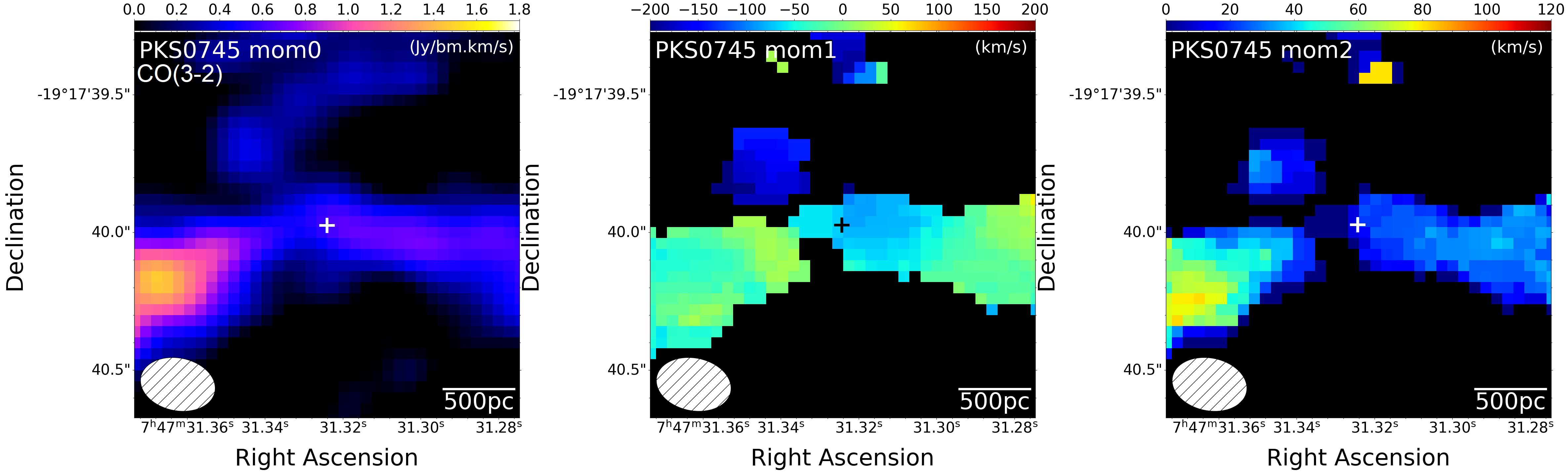}
 \end{center}
\caption{Continued}
\end{figure}

\begin{figure}[htbp]
 \centering
\begin{minipage}[t]{0.30\textwidth}
 \includegraphics[width=5.0cm]{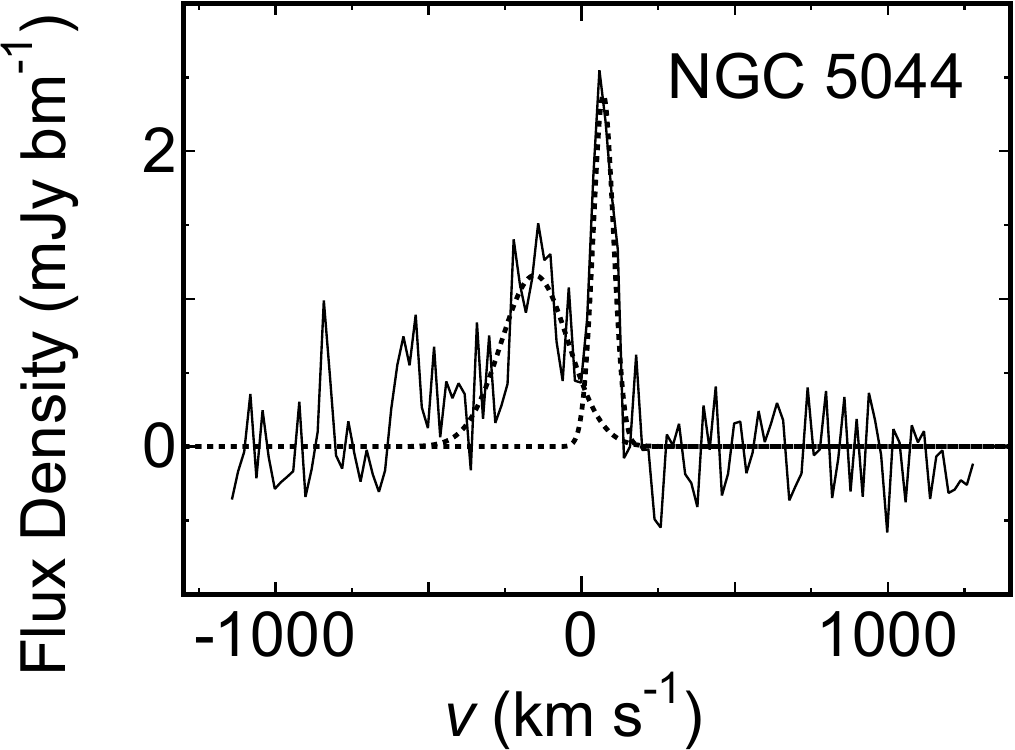}
\end{minipage}
\begin{minipage}[t]{0.30\textwidth}
 \includegraphics[width=5.0cm]{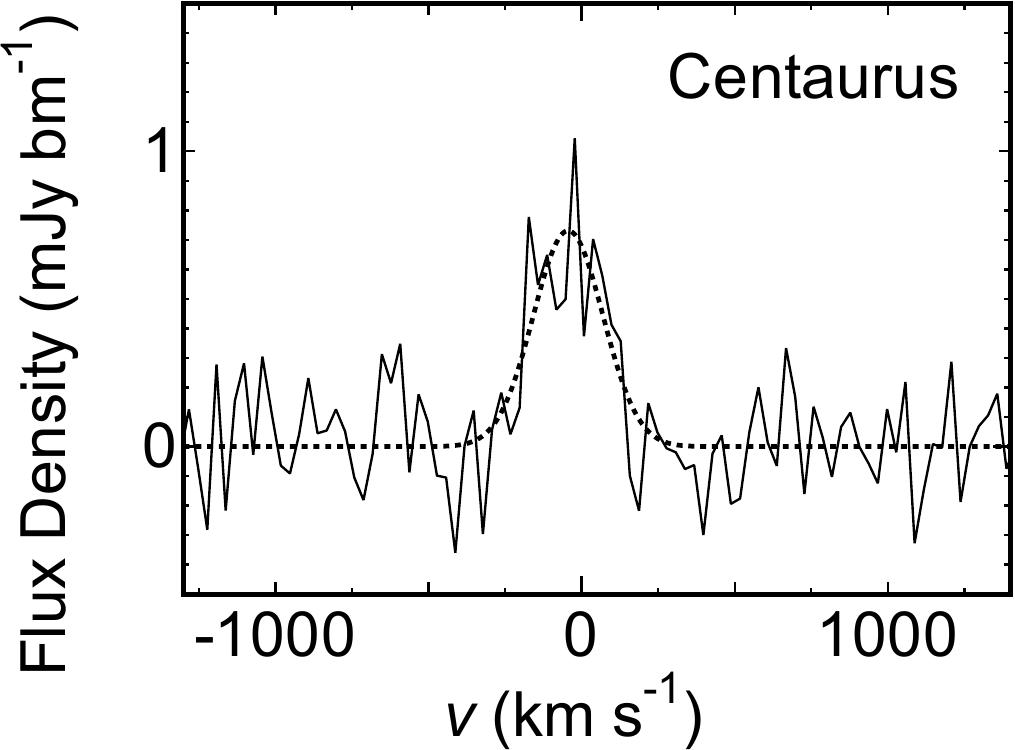}
\end{minipage}
\begin{minipage}[t]{0.30\textwidth}
 \includegraphics[width=5.0cm]{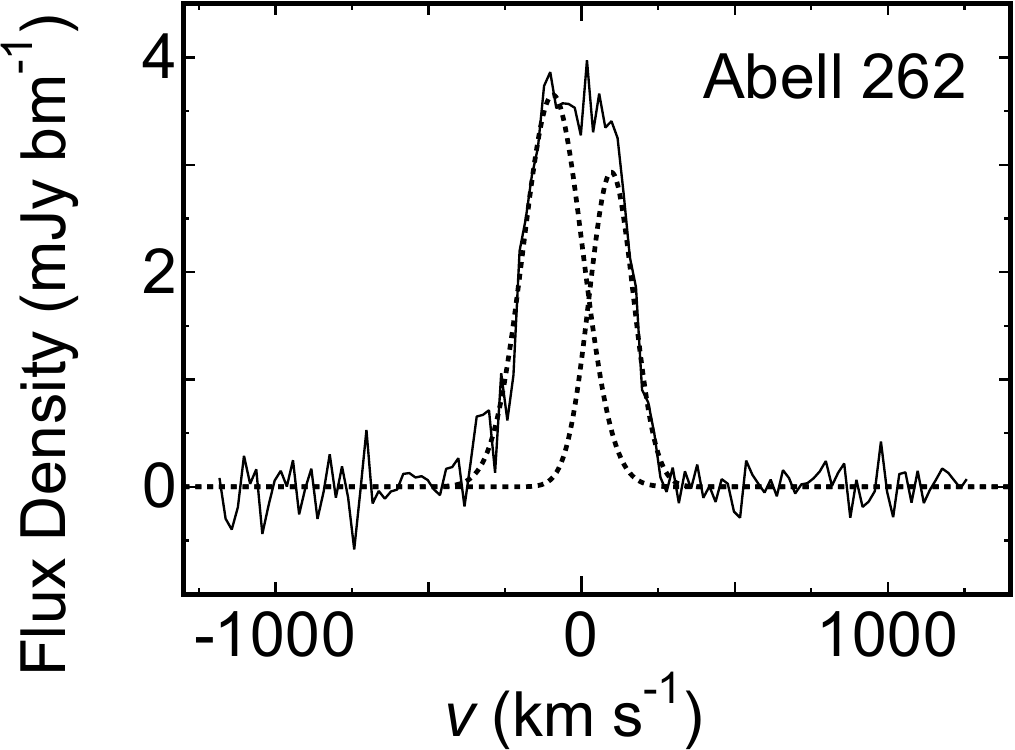}
\end{minipage} \\
\begin{minipage}[t]{0.30\textwidth}
 \includegraphics[width=5.0cm]{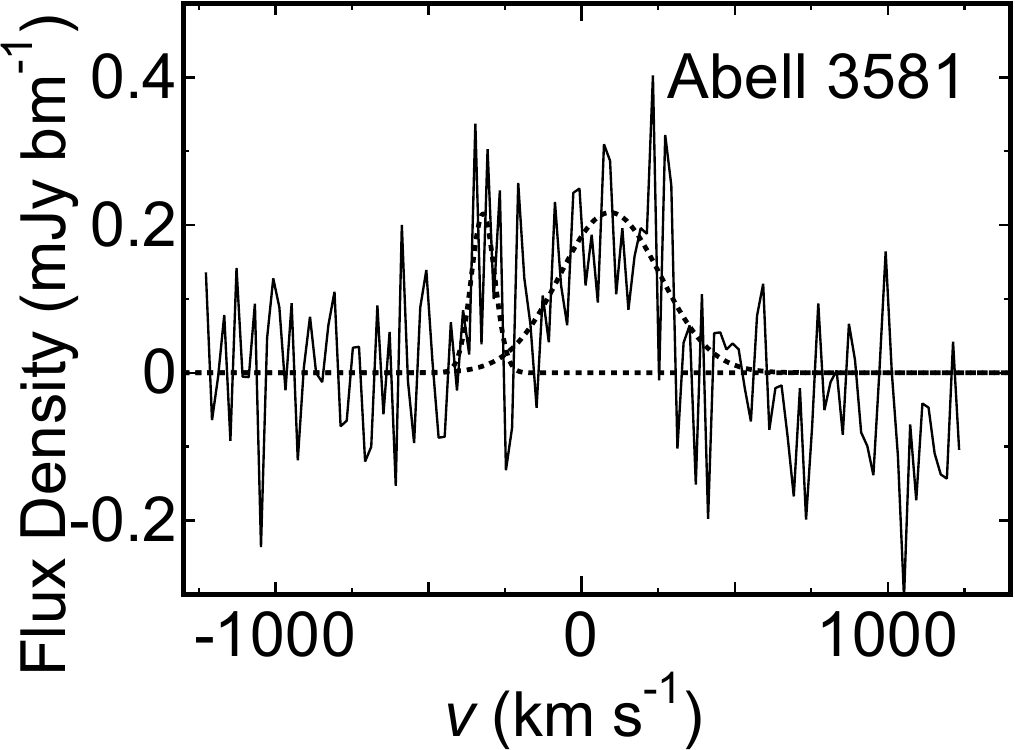}
\end{minipage}
\begin{minipage}[t]{0.30\textwidth}
 \includegraphics[width=5.0cm]{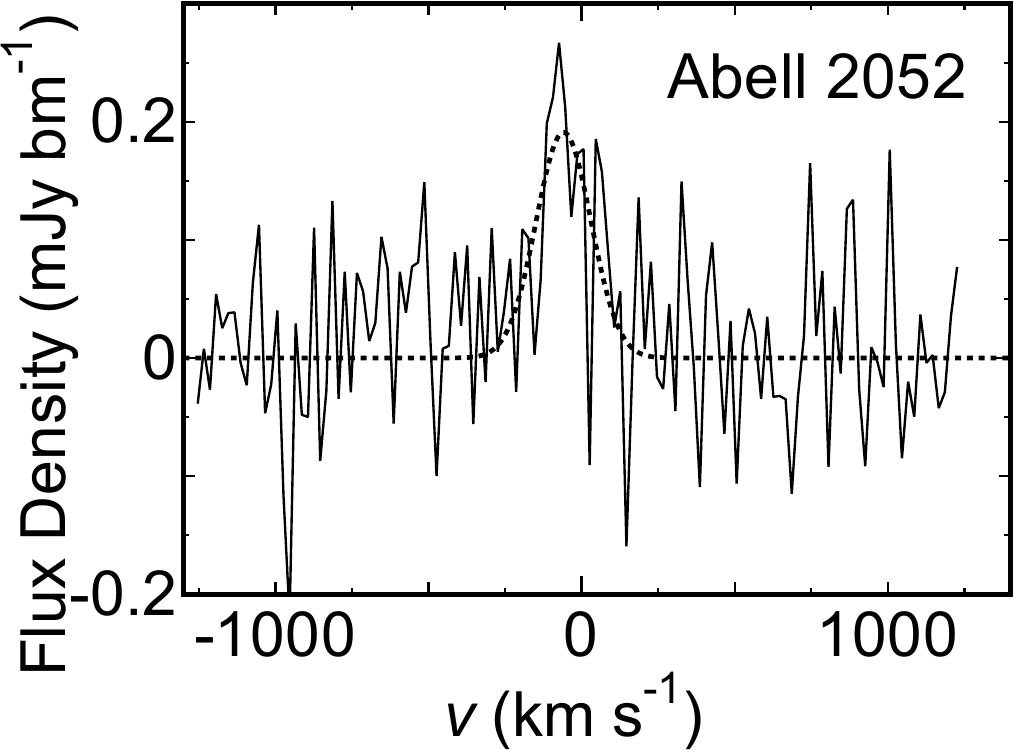}
\end{minipage}
\begin{minipage}[t]{0.30\textwidth}
 \includegraphics[width=5.0cm]{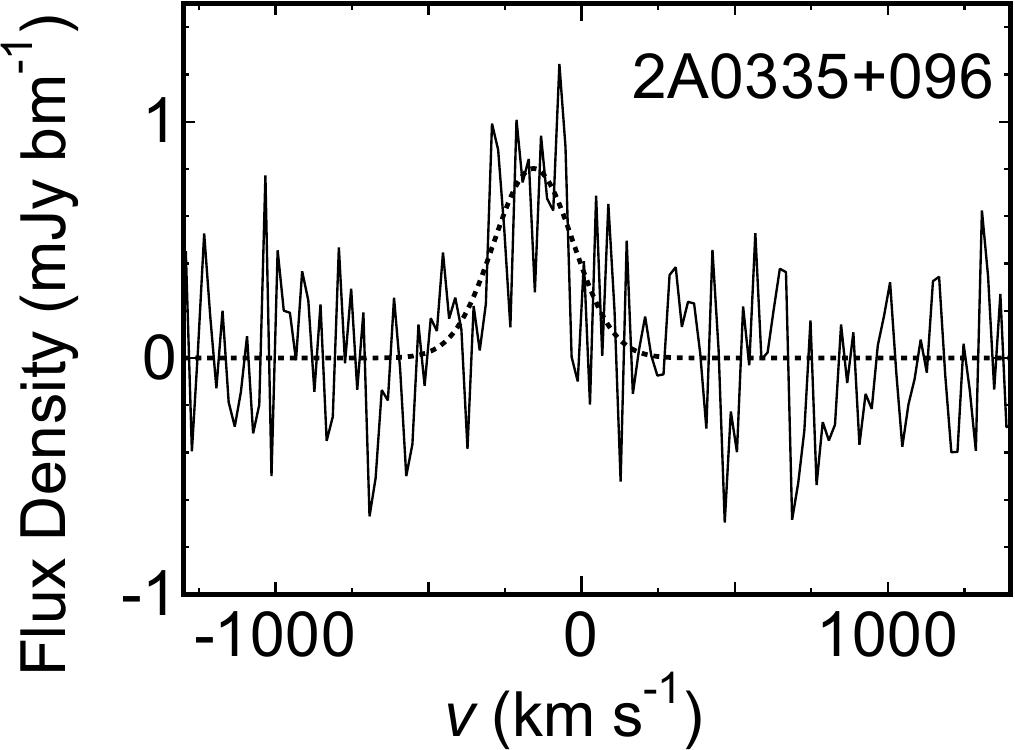}
\end{minipage} \\
\begin{minipage}[t]{0.30\textwidth}
 \includegraphics[width=5.0cm]{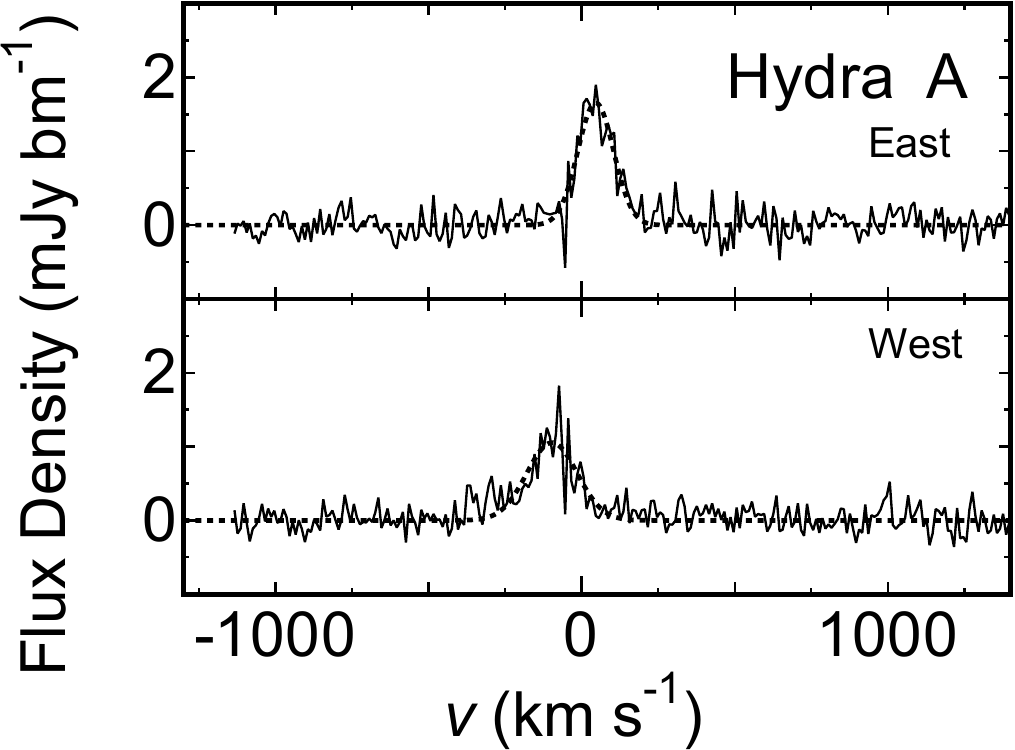}
\end{minipage}
\begin{minipage}[t]{0.30\textwidth}
 \includegraphics[width=5.0cm]{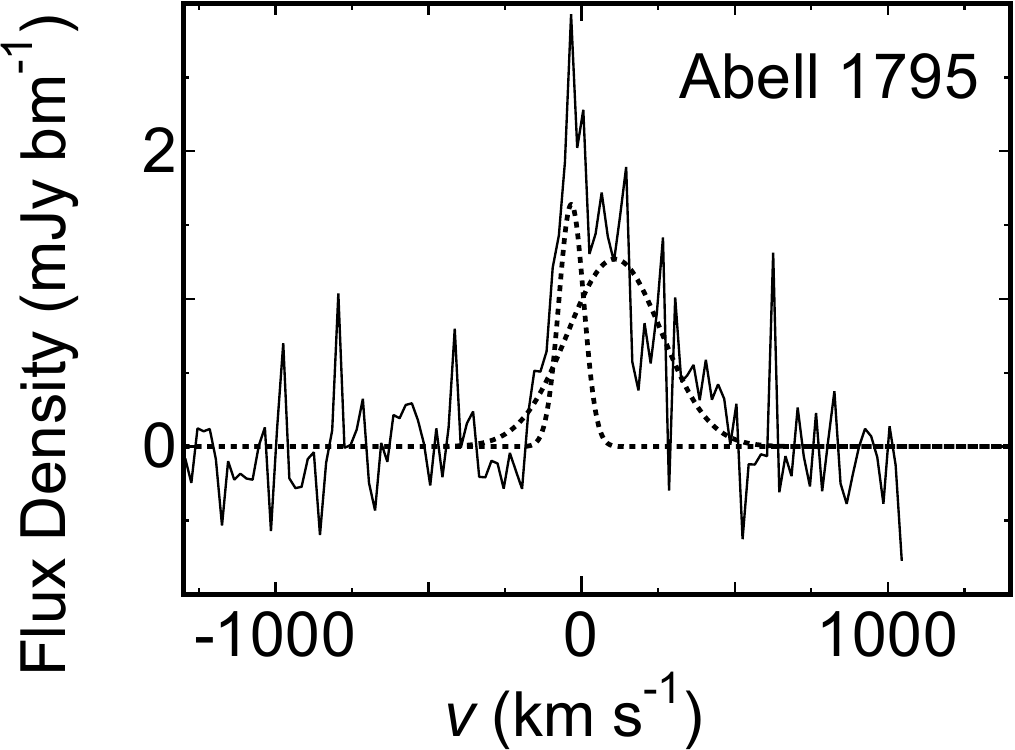}
\end{minipage}
\begin{minipage}[t]{0.30\textwidth}
 \includegraphics[width=5.0cm]{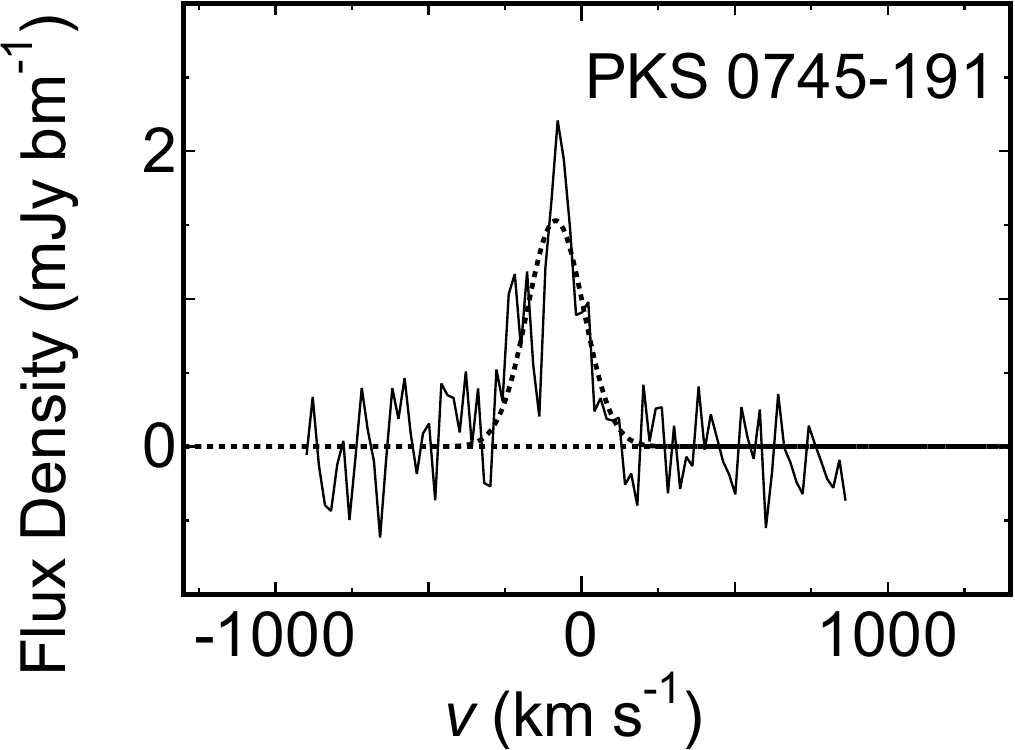}
\end{minipage} 
 \caption{CO line spectrum within 500~pc from the AGNs. The
dotted lines show the best-fit gaussian components.}\label{fig:spec}
\end{figure}

\begin{table}
  \tbl{Observed CO and mm/submm continuum emissions.}{%
  \begin{tabular}{ccccccc}
      \hline
Target & $v$\footnotemark[$*$] & FWHM\footnotemark[$\dag$] & $S_{\rm CO}\Delta v$ & $\nu_{\rm con}$ & $F_{\rm con}$ & $\beta_{\rm con}$\footnotemark[$\ddag$]  \\
 & ($\rm km\: s^{-1}$) & ($\rm km\: s^{-1}$) & ($\rm mJy\:  bm^{-1}\: km\: s^{-1}$) & (GHz) & (mJy) &  \\
      \hline
NGC 5044 & $70.8 \pm 3.9$ & $77.7 \pm 9.9$ & $197 \pm 32$ & 235.2 & $49.5\pm 3.7$ & $0.22\pm 2.46$ \\
 & $-153 \pm 16$ & $259 \pm 42$ & $320 \pm 63$ &  &  &    \\
 &  & ($190\pm 34$) &  &  &  &    \\
Centaurus & $-44 \pm 15$ & $268 \pm 36$ & $208 \pm 37$ & 107.1 & $51.6\pm 2.6$ & $-1.30\pm 0.49$ \\
Abell  262 & $96.9 \pm 9.1$ & $170 \pm 14$ & $529 \pm 67$ & 235.6 & $3.336\pm 0.069$ & $-0.76\pm 1.13$ \\
 & $-94 \pm 11$ & $229 \pm 18$ & $890 \pm 76$ &  &  &    \\
 &  & ($207\pm 17$) &  &  &  &    \\
Abell 3581 & $-322 \pm 15$ & $90 \pm 36$ & $21 \pm 11$ & 231.8 & $107.5\pm 3.7$ & $-0.22\pm 0.77$ \\
 & $93 \pm 31$ & $388 \pm 76$ & $90 \pm 23$ &  &  &    \\
 &  & ($332\pm 70$) &  &  &  &    \\
Abell  2052 & $-59 \pm 17$ & $211 \pm 41$ & $43 \pm 11$ & 229.5 & $34.3\pm 3.1$ & $-1.08\pm 1.29$ \\
2A0335+096 & $-157 \pm 24$ & $307 \pm 56$ & $262 \pm 63$ & 110.5 & $6.01\pm 0.34$ & ... \\
Hydra A & $-96.8 \pm 5.6$ & $194 \pm 13$ & $217 \pm 20$ & 227.6 & $65.23\pm 0.19$ & $-0.749\pm  0.062$ \\
 & $46.0 \pm 3.0$ & $137.7 \pm 7.1$ & $243 \pm 16$ &  &  &    \\
 &  & ($164\pm 10$) &  &  &  &    \\
Abell  1795 & $-33.7 \pm 6.8$ & $95 \pm 20$ & $166 \pm 46$ & 225.8 & $3.29\pm 0.041$ & $-2.06\pm 0.41$ \\
 & $108 \pm 27$ & $362 \pm 44$ & $488 \pm 81$ &  &  &    \\
 &  & ($294\pm 39$) &  &  &  &    \\
PKS 0745-191 & $-85.2 \pm 9.7$ & $217 \pm 23$ & $353 \pm 49$ & 314.6 & $5.03\pm 0.35$ & ... \\
      \hline
    \end{tabular}}\label{tab:result}
\begin{tabnote}
\footnotemark[$*$] The velocity of the line component relative to the systemic velocity of the galaxy.\\
\footnotemark[$\dag$] Flux-averaged ones are parenthesized.  \\
\footnotemark[$\ddag$]  The spectral index of the continuum
emission around $\nu_{\rm con}$, which cannot be constrained for 2A0335+096 and PKS~0745-191 (see text).
\end{tabnote}
\end{table}

\subsection{Molecular gas mass}

From the CO intensities $S_{\rm CO}\Delta v$ presented in
table~\ref{tab:result}, we derive the masses of molecular gas within
$500$~pc from the AGNs using the following relation from
\citet{2013ARA&A..51..207B}:
\begin{equation}
 M_{\rm mol} = \frac{1.05\times 10^4}{F_{\rm ul}}\left(\frac{X_{\rm CO}}{2\times 10^{20}
\frac{\rm cm^{-2}}{\rm K\: km\: s^{-1}}}\right)
\left(\frac{1}{1+z}\right)
\left(\frac{S_{\rm CO}\Delta v}{\rm Jy\: km\: s^{-1}}\right)
\left(\frac{D_{\rm L}}{\rm Mpc}\right)^2\: M_\odot\:,
\end{equation}
where $X_{\rm CO}$ is the CO-to-H$_2$ conversion factor, and $D_{\rm L}$ is
the luminosity distance. 
The factor $F_{\rm ul}$ represents the assumption on the CO excitation, which is necessary to estimate CO(J=1-0) flux from higher-J measurements. Specifically, $F_{21}$ gives a CO(J=2-1)/CO(J=1-0) line ratio in flux density scale, 3.2, and $F_{32}$ is a CO(J=3-2)/CO(J=1-0) ratio of 7.2
\citep{2019MNRAS.485..229R,2019MNRAS.490.3025R}. The dispersion of the values among the galaxies is about $\sim 40$~\% \citep{1992A&A...264..433B,2003A&A...412..657S}. The conversion factor
$X_{\rm CO}$ is not well-understood for elliptical galaxies. Thus, we
adopt a standard value of $X_{\rm CO}=2\times 10^{20}\rm\: cm^{-2}(K\:
km\: s^{-1})^{-1}$ measured in the Milky Way, following previous
studies (\cite{2019A&A...631A..22O,2019MNRAS.490.3025R} and see their
discussion). The results are shown in table~\ref{tab:Mmol}. When there
are two gaussian components in the spectrum, the mass is their
summation.

\begin{table}
  \tbl{Molecular gas mass and mm/submm continuum luminosity obtained in
  this study. }{
  \begin{tabular}{ccccc}
      \hline
Target & $M_{\rm mol}$ & $t_{\rm acc}$ & $\dot{M}$ & $L_{\rm con}$\footnotemark[$*$] \\
 & ($10^7\: M_\odot$) & ($10^7$~yr) & ($M_\odot\rm\: yr^{-1}$) & $(10^{42}\rm\: erg\: s^{-1})$ \\
      \hline
NGC 5044 & $2.0\pm 0.2$ & 4.0 & $0.48\pm 0.13$ & $(1.9\pm 0.1)\times 10^{-2}$ \\
Centaurus & $1.1\pm 0.2$ & 2.9 & $0.38\pm 0.12$ & $(1.2\pm 0.1)\times 10^{-2}$ \\
Abell  262 & $24\pm 2$ & 3.7 & $6.6\pm 1.0$ & $(3.9\pm 0.1)\times 10^{-3}$ \\
Abell 3581 & $2.9\pm 0.7$ & 2.3 & $1.2\pm 0.5$ & $0.23\pm 0.01$ \\
Abell  2052 & $5.4\pm 1.5$ & 3.6 & $1.5\pm 0.7$ & $0.18\pm 0.02$ \\
2A0335+096 & $6.1\pm 1.5$ & 2.5 & $2.4\pm 1.0$ & $(2.0\pm 0.1)\times 10^{-2}$ \\
Hydra A & $42 \pm 2$ & 4.7 & $8.9\pm 1.0$ & $0.87\pm 0.0$ \\
Abell  1795 & $15\pm 2$ & 2.6 & $5.8\pm 2.0$ & $(6.0\pm 0.1)\times 10^{-2}$ \\
PKS 0745-191 & $39\pm 5$ & 3.5 & $11\pm 3$ & $0.32\pm 0.02$ \\
      \hline
    \end{tabular}}\label{tab:Mmol}
\begin{tabnote}
\footnotemark[$*$] Calculated from $\nu_{\rm con}$ and $F_{\rm con}$ shown in table~\ref{tab:result} using equation~(\ref{fig:dMdt-Pcav}).  \\ 
\end{tabnote}
\end{table}

\begin{table}
  \tbl{Properties of the targets. }{%
  \begin{tabular}{ccccc}
      \hline
Target & $P_{\rm cav}$\footnotemark[$*$] & $F_{1.4}$\footnotemark[$\dag$] & $L_{\rm 1.4}$ & $M_{\rm BH}$\footnotemark[$\ddag$] \\
 & ($10^{42}\rm\: erg^{-1}$) & (mJy) & ($10^{42}\rm\: erg^{-1}$) & ($10^9\: M_\odot$) \\
      \hline
NGC 5044 & $4.2_{-2.0}^{+1.2}$ & $34.7\pm 1.1$ & $(9.3\pm 0.3)\times 10^{-5}$ & $0.2$ \\
Centaurus & $7.4_{-1.8}^{+5.8}$ & $3980\pm 110$ & $(1.2\pm 0.0)\times 10^{-2}$ & $0.4_{-0.1}^{+0.1}$ \\
Abell  262 & $9.7_{-2.6}^{+7.5}$ & $65.7\pm 2.3$ & $(5.4\pm 0.2)\times 10^{-4}$ & $0.4_{-0.1}^{+0.2}$ \\
Abell 3581 & $3.1$ & $645.5\pm 22.8$ & $(9.7\pm 0.3)\times 10^{-3}$ & $0.4_{-0.1}^{+0.1}$ \\
Abell  2052 & $150_{-70}^{+200}$ & $5499\pm 209$ & $(2.1\pm 0.1)\times 10^{-1}$ & $0.4_{-0.1}^{+0.2}$ \\
2A0335+096 & $24_{-6}^{+23}$ & $36.7\pm 1.8$ & $(1.6\pm 0.1)\times 10^{-3}$ & $0.5_{-0.2}^{+0.3}$ \\
Hydra A & $430_{-50}^{+200}$ & $40850 \pm 1279$ & $4.0\pm 0.1$ & $0.9_{-0.4}^{+0.7}$ \\
Abell  1795 & $160_{-50}^{+230}$ & $925\pm 28$ & $(1.2\pm 0.0)\times 10^{-1}$ & $0.6_{-0.2}^{+0.3}$ \\
PKS 0745-191 & $1700_{-300}^{+1400}$ & $2372\pm 84$ & $(8.8\pm 0.3)\times 10^{-1}$ & $1.1_{-0.4}^{+0.7}$ \\
      \hline
    \end{tabular}}\label{tab:pro}
\begin{tabnote}
\footnotemark[$*$] Jet power from \citet{2006ApJ...652..216R} and \citet{2018ApJ...853..177P}. \\
\footnotemark[$\dag$] 1.4~GHz flux from \citet{1998AJ....115.1693C} except for Centaurus \citep{1981A&AS...45..367K} \\ 
\footnotemark[$\ddag$] Black hole mass from \citet{2006ApJ...652..216R} except for NGC~5044 \citep{2009ApJ...705..624D}.\\ 
\end{tabnote}
\end{table}

\subsection{Circumnuclear gas and jet power}
\label{sec:cgas}

\begin{figure}
 \begin{center}
  \includegraphics[width=8cm]{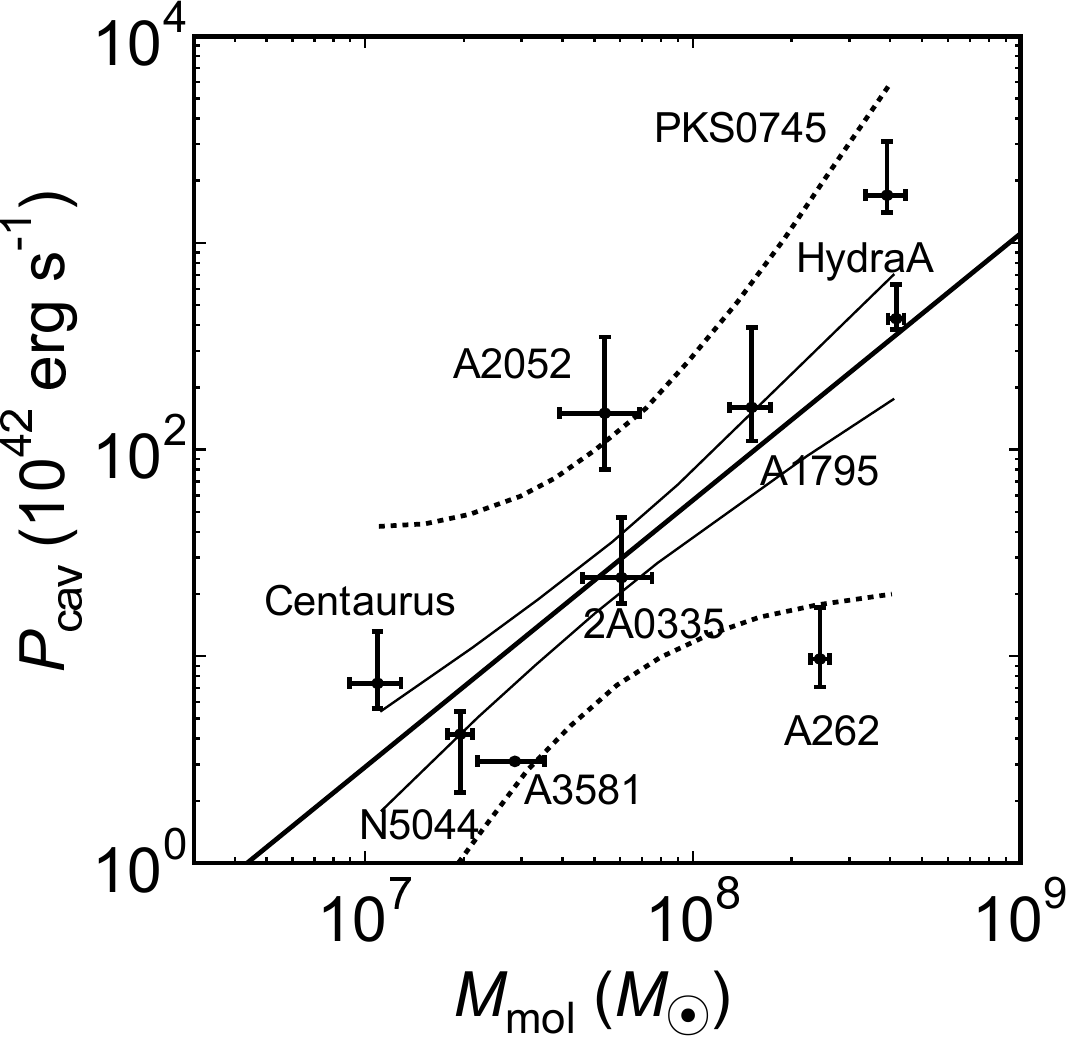} 
 \end{center}
\caption{Molecular gas mass within 500~pc from the AGN compared with the
jet power estimated from the X-ray cavities. The thick solid line shows
the best-fitting power-law model, determined using the BCES
estimator. The thin solid and dotted lines show the $1\:\sigma$ and
$3\:\sigma$ confidence ranges, respectively.}\label{fig:M-Pcav}
\end{figure}

Figure~\ref{fig:M-Pcav} shows the relation between the molecular gas
mass at $r<500$~pc ($M_{\rm mol}$) and the jet power required to inflate
X-ray cavities ($P_{\rm cav}$). The data of $P_{\rm cav}$ have been
taken from the literature
\citep{2006ApJ...652..216R,2018ApJ...853..177P}. A correlation can be
seen between $M_{\rm mol}$ and $P_{\rm cav}$, and the Pearson
correlation coefficient on the logarithmic scales is 0.754 with
$p$-value 0.019. The correlation can be described by a power-law model
of the form:
\begin{equation}
\label{eq:A1}
 \log\left(\frac{P_{\rm cav}}{10^{42}\rm erg\: s^{-1}}\right) = A_1\log\left(\frac{M_{\rm mol}}{M_\odot}\right) + B_1\:.
\end{equation}
Using the bivariate correlated errors and intrinsic scatter
BCES $(Y|X)$ estimator \citep{1996ApJ...470..706A}, which accounts for
errors in both axes and the presence of possible intrinsic scatter, we
obtain $A_1 = 1.3\pm 0.4$ and $B_1 = -8.6\pm 3.0$. In other words, the
correlation indicates that the jet power is proportional to the average
density of the circumnuclear molecular gas because we considered the
mass within the fixed radius and $A_1\sim 1$. Abell~262 appears to be an
outlier of this correlation because it is outside the $3\:\sigma$
confidence range (figure~\ref{fig:M-Pcav}). If we exclude Abell~262 from
the sample, the correlation coefficient increases to 0.906 with
$p$-value 0.0020. Abell~262 has a lower AGN activity despite retaining a
large amount of gas. Possible reasons will be discussed in
section~\ref{sec:circ}.

The gas accretion rate $\dot{M}$ may be more directly related to the AGN
activity than $M_{\rm mol}$. The accretion is likely to be induced by
the angular momentum transfer by turbulence
\citep{2010MNRAS.408..961P,2017MNRAS.466..677G}. In this case, the
accretion process can be regarded as a diffusion process of gas clumps,
and the diffusion coefficient is $\nu \sim \ell_{\rm t} v_{\rm t}$,
where $\ell_{\rm t}$ and $v_{\rm t}$ is the eddy scale and the velocity
of the turbulence, respectively 
(e.g. \cite{1995ApJ...440..674X}) Thus, the diffusion or the
accretion time scale of the gas on the scale of interest ($r_{\rm g}$)
is $t_{\rm acc}\sim r_{\rm g}^2/\nu$, and the accretion rate is
$\dot{M}\sim M_{\rm mol}/t_{\rm acc}$. We note that this estimation can
even be applied to the standard disk \citep{1973A&A....24..337S}. Since
the kinetic viscosity coefficient can be regarded as the diffusion
coefficient, the accretion time of a disk on a scale of $r_{\rm g}$ is
also $t_{\rm acc}\sim r_{\rm g}^2/\nu$ (section 6 of
\cite{1981ARA&A..19..137P}). Thus, we adopt the relation of $\dot{M}=
M_{\rm mol}/t_{\rm acc}$ regardless of whether molecular gas has a
disk-like structure or not. We assume that $v_t={\rm FWHM}/(2\sqrt{2\ln
2})\approx {\rm FWHM}/2.35$ and $\ell_{\rm t}=\alpha r_{\rm g}$
\citep{1983bhwd.book.....S}, where $r_{\rm g}=500$~pc and $\alpha =
0.15$.  For the FWHM, we use the flux averaged values when there are two
gaussian components (parenthesized ones in table~\ref{tab:Mmol}).  We
note that the derived $v_t$ may include a rotational component on a
small scale that cannot be spatially resolved (see
\cite{2019ApJ...883..193N}) and it should be regarded as the upper
limit. However, the effect of rotation velocity contamination
may not be significant. We have investigated the effect within the
observing beam, using a cube model code 3D-Barolo
\footnote{https://editeodoro.github.io/Bbarolo/} for Abell~262, a galaxy
with a spatially resolved CO velocity field. The model fit shows that
$v_t\sim 100\rm\: km\: s^{-1}$, which is close to FWHM/2.35$\sim 90\rm\:
km\: s^{-1}$ (table~\ref{tab:result}).  We also indicate that not all
the in-flowing gas ($\dot{M}$) is eventually swallowed by the black hole
because some of it is consumed in star formation
\citep{2022ApJ...924...24F}. The obtained $t_{\rm acc}$ and $\dot{M}$
are shown in table~\ref{tab:Mmol}.

\begin{figure}
 \begin{center}
  \includegraphics[width=8cm]{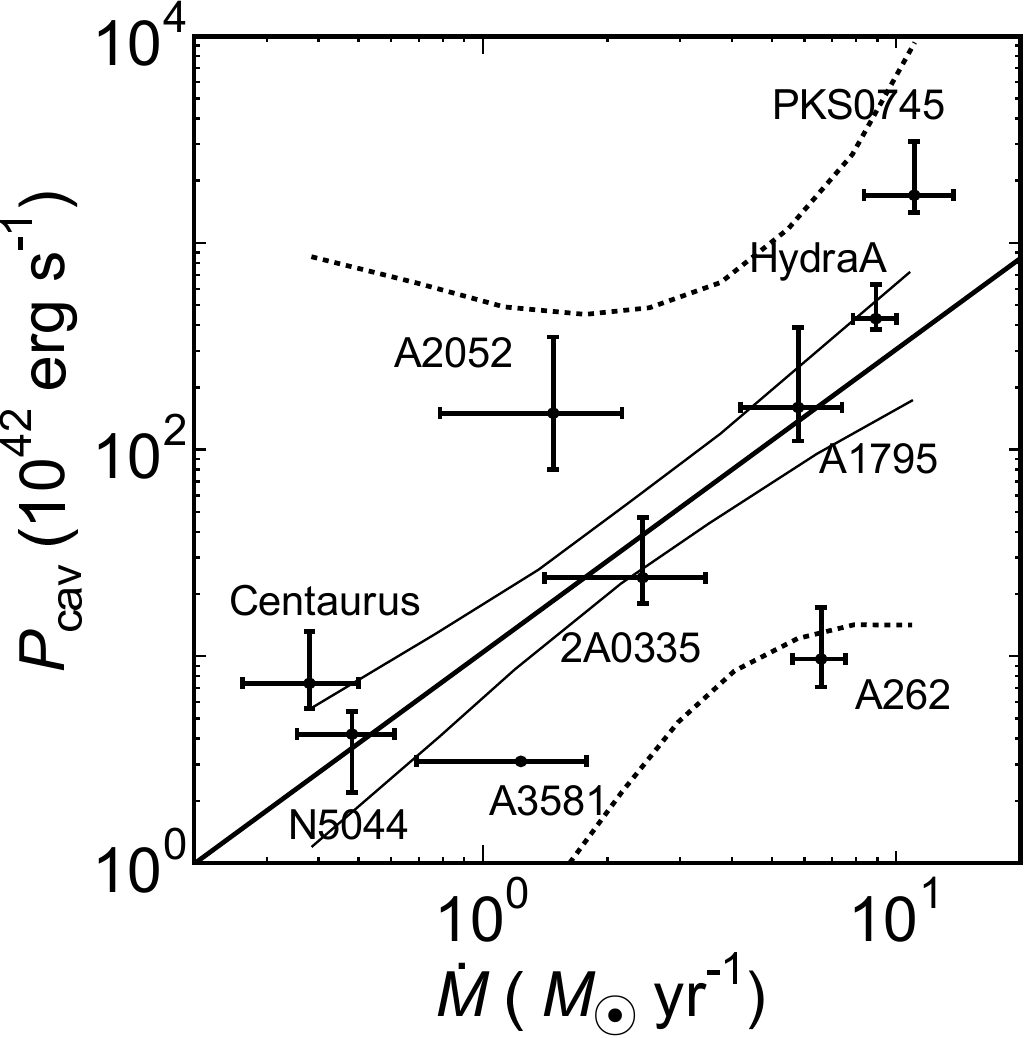} 
 \end{center}
\caption{Same as figure~\ref{fig:M-Pcav} but the gas accretion rate
compared with the jet power estimated from X-ray cavity
observations.}\label{fig:dMdt-Pcav}
\end{figure}

Figure~\ref{fig:dMdt-Pcav} shows the relation between $\dot{M}$ and
$P_{\rm cav}$, which is similar to that between $M_{\rm mol}$ and
$P_{\rm cav}$ (figure~\ref{fig:M-Pcav}).  The correlation coefficient is
0.733 with $p$-value 0.025, and it is 0.871 with $p$-value 0.0049 if we
exclude Abell~262 because it is almost at the end of the $3\:\sigma$
confidence range. The correlation among the 9 objects can be represented
by a power-law model of the form:
\begin{equation}
 \log\left(\frac{P_{\rm cav}}{10^{42}\rm erg\: s^{-1}}\right) = A_2\log\left(\frac{\dot{M}}{M_\odot\:\rm yr^{-1}}\right) + B_2\:.
\end{equation}
The BCES $(Y|X)$ estimator gives $A_2=1.5\pm 0.6$ and $B_2=1.0\pm 0.3$.
The similarity between figures~\ref{fig:M-Pcav} and~\ref{fig:dMdt-Pcav}
reflects the fact that while FWHM varies only a factor of few, $M_{\rm
mol}$ ranges nearly two orders of magnitude. This suggests that the mass
of circumnuclear molecular gas is the main factor that determines the
jet activity. Thus, the $\dot{M}$--$P_{\rm cav}$ relation seems to be
robust even if we consider potential uncertainties such as the gas
morphology (disk-like or not) and the contamination of a rotational
velocity component.

\subsection{Circumnuclear gas and continuum emission}
\label{sec:MLcon}

Since radio continuum emission is observed as a point source at the
position of the AGNs, it should come from the vicinity of the AGNs
($\lesssim 500$~pc) and reflect their recent activity. We derive the
continuum flux at line-free channels. 

\begin{figure}
 \begin{center}
  \includegraphics[width=8cm]{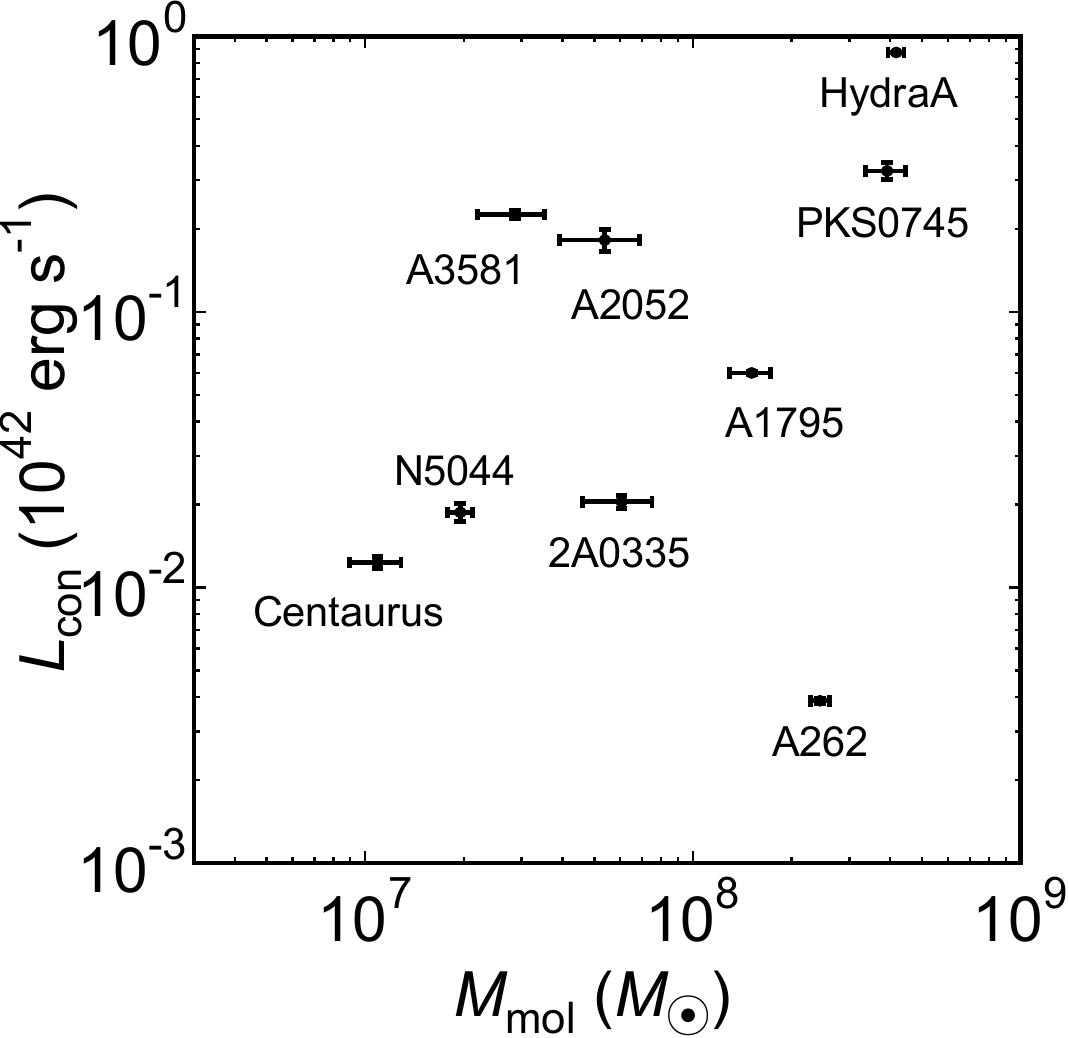} 
 \end{center}
\caption{Continuum luminosity in the mm/submm band compared with the
molecular gas mass within 500~pc from the AGN.}\label{fig:M-Lcon}
\end{figure}

We convert fluxes at observed frequencies to those at the rest-frame
CO(J=1-0) line frequency ($\nu_{10}=115.3$~GHz) assuming that the
continuum emission is represented by a power-law $S_\nu \propto
(\nu/\nu_{10})^{\gamma}$, where $\nu$ is the frequency. We adopt a
typical spectral index of $\gamma=-0.75$. For an object at the redshift
$z$ and observed at the frequency $\nu_{\rm con}$, the radio continuum
luminosity at the frequency $\nu_{10}$ is estimated to be
\begin{equation}
 L_{\rm con} = \frac{4\pi D_{\rm L}^2 \nu_{\rm con} F_{\rm con}}{[(1+z)\eta]^{1+\gamma}}\:,
\end{equation}
where $F_{\rm con}$ is the observed flux (mJy) and $\eta$ is the line
frequency ratios ($\eta=1$ for CO(J=1-0), $\eta\approx 2$ for
CO(J=2-1)\footnote{Exactly speaking, the continuum flux is not
measured right at the frequency of the CO line. However, since the index
is $1+\gamma=0.35$, the minor modification $\eta$ does not affect the
results.}, and $\eta\approx 3$ for CO(J=3-2)), which is required for the
flux conversion among different line frequencies. The derived
luminosities $L_{\rm con}$ are presented in table~\ref{tab:Mmol} and
they are compared with $M_{\rm mol}$ in figure~\ref{fig:M-Lcon}. The
correlation coefficient is 0.386 and the $p$-value is 0.30, which is
much larger than the conventional threshold of 0.05. Thus, it is
unlikely that that $L_{\rm con}$ is correlated with $M_{\rm mol}$.
Again, Abell~262 appears to be an outlier because the radio luminosity
is extremely small for its large $M_{\rm mol}$. If we exclude Abell~262
from the sample, the coefficient is 0.740 with $p$-value 0.036, which
shows a sign of correlation.

\begin{figure}
 \begin{center}
  \includegraphics[width=8cm]{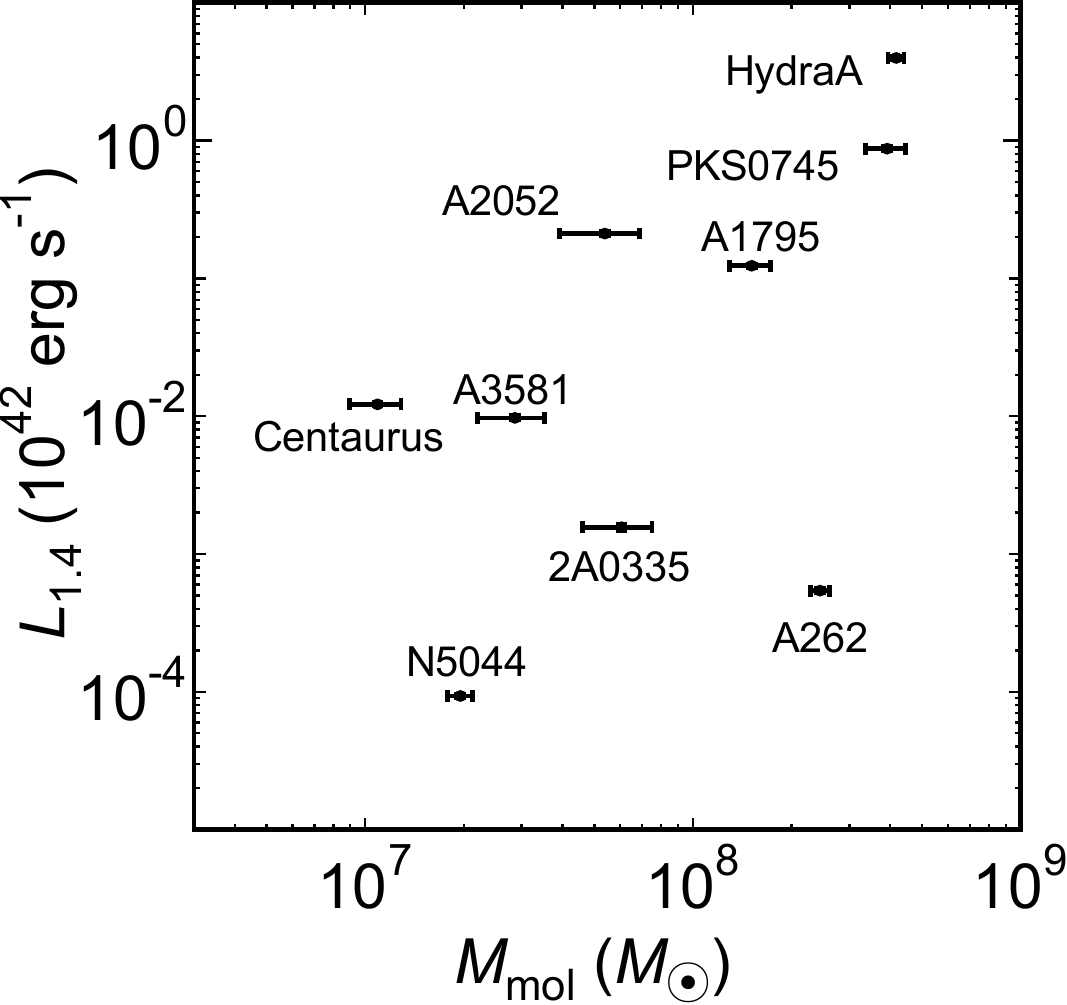} 
 \end{center}
\caption{Continuum luminosity at $1.4$~GHz compared with the
molecular gas mass within 500~pc from the AGN .}\label{fig:M-L14}
\end{figure}

Traditionally, radio emissions at lower frequencies have been used as an
indicator of AGN activities
\citep{2004ApJ...607..800B,2018ApJ...853..177P}. Figure~\ref{fig:M-L14}
compares the radio luminosity estimated from 1.4~GHz observations
(NRAO/VLA Sky Survey\footnote{https://www.cv.nrao.edu/nvss/NVSSlist.shtml}; \cite{1998AJ....115.1693C}) with the mass of the
circumnuclear molecular gas ($M_{\rm mol}$). The former is derived by
\begin{equation}
 L_{\rm 1.4} = \frac{4\pi D_{\rm L}^2 \nu_{\rm 1.4} F_{\rm 1.4}}
{(1+z)^{1+\gamma}}\:,
\end{equation}
where $\nu_{1.4}$ is $1.4$~GHz, $F_{\rm 1.4}$ is the observed flux at
1.4~GHz, and $\gamma=-0.75$. Since our targets are at lower redshifts
($z\lesssim 0.1$), the correction of $(1+z)^{1+\gamma}$ is insensitive to
$\gamma$. The correlation coefficient is 0.556 with $p$-value 0.12. Thus,
$L_{\rm 1.4}$ is unlikely to be correlated with $M_{\rm mol}$. However,
if Abell~262 is ignored, the correlation coefficient is 0.778 with
$p$-value 0.023, which may indicate the existence of correlation.

\section{Discussion}

\subsection{Circumnuclear gas and AGN activity}
\label{sec:circ}

We have shown that the mass of circumnuclear molecular gas ($M_{\rm
mol}$) within a fixed radius (500~pc) has a correlation with the jet
power estimated from X-ray cavities ($P_{\rm cav}$;
figure~\ref{fig:M-Pcav}). This suggests that the circumnuclear molecular
gas serves as the fuel for AGN feedback
\citep{2022ApJ...924...24F}. Assuming that the gas accretion is caused
by angular momentum transport by turbulence, the accretion rate
($\dot{M}$) is also correlated with $P_{\rm cav}$
(figure~\ref{fig:dMdt-Pcav}), although the $\dot{M}$--$P_{\rm cav}$
relation is virtually the $M_{\rm mol}$--$P_{\rm cav}$ relation. We
conclude that the jet power depends on the average density of the
circumnuclear molecular gas because we considered the mass within the
fixed radius. Since $A_1 = 1.3\pm 0.4$ in equation~(\ref{eq:A1}), a
simple relation of $P_{\rm cav}\propto M_{\rm mol}$ is acceptable. This
fact would be useful when constructing sub-grid models of AGN feedback
in numerical simulations, which have often adopted the classical Bondi
accretion rate. The correlation between the mass of circumnuclear
molecular gas and the AGN activity has been known for disk galaxies
(e.g. \cite{2016ApJ...827...81I}, but there is a counter-example, see \cite{2020ApJ...895...73K}). Thus, the AGN fueling mechanism for
elliptical galaxies may be similar to that in the disk galaxies.

In contrast with $P_{\rm cav}$, the continuum luminosities both at $\sim
1.4$~GHz ($L_{1.4}$) and $\sim 100$--300~GHz ($L_{\rm col}$) seem to
have no correlation with $M_{\rm mol}$ (figures~\ref{fig:M-Lcon}
and~\ref{fig:M-L14}). The difference may be ascribed to the relevant
time scales of AGN activities. The jet power $P_{\rm cav}$ is often
described by the ratio of the energy required to inflate the cavity to
its evolution time ($t_{\rm cav}$) such as the cavity's buoyancy age
(\cite{2006MNRAS.372...21A,2006ApJ...652..216R}; see also
\cite{2016PASJ...68...26F}). Thus, $P_{\rm cav}$ reflects the AGN
activity averaged over on a time scale of $t_{\rm cav}$, which is
estimated to be $\sim 10^7$~yr
\citep{2006ApJ...652..216R}. Interestingly, it is comparable to the gas
accretion time $t_{\rm acc}$ (table~\ref{tab:Mmol}). On the other hand,
since the mm/submm ($\sim 100$--300~GHz) continuum is observed as nuclear
point sources coincident with the AGNs, the size of the emitting regions
should be $\lesssim 500$~pc. Since the light-travel time of the regions
is only $\lesssim 1600$~yr, the emission virtually reflects the current
activity of the AGN. The existence of the correlation between $P_{\rm
cav}$ and $M_{\rm mol}$ and the non-existence between the continuum
luminosities and $M_{\rm mol}$ suggest that the circumnuclear gas is
rather persistently fueling the AGN activity on a longer time scale
($t_{\rm cav}\sim 10^7$~yr), while the continuum
luminosities vary on a short time scale. However, since $L_{\rm 1.4}$ is the total
luminosity of the galaxy, it may include emissions from larger and older
structures such as $\sim$~kpc radio jets and lobes\footnote{The angular resolution of the observations is $45''$ \citep{1998AJ....115.1693C}, which corresponds to 18~kpc at $z=0.02$.}. 
Thus, 
it could be proven that $L_{\rm 1.4}$ is more strongly correlated with $M_{\rm mol}$ than $L_{\rm con}$, once a larger sample becomes available.

In figures~\ref{fig:M-Pcav}--\ref{fig:M-L14}, Abell~262 appears to be an
outlier. While it has a large amount of circumnuclear molecular gas
($M_{\rm mol}$), both the jet power ($P_{\rm cav}$) and the 
continuum luminosities ($L_{\rm con}$ and $L_{\rm 1.4}$) are small.
This might be explained if the AGN activity has been gentle for a long
time. Thus, there may be an additional factor that affects the activity
other than the mass of circumnuclear molecular gas. For example,
compared with other galaxies with similar amount of $M_{\rm mol}$
(Abell~1795, Hydra~A, and PKS~0745--191), the mass of the black hole
($M_{\rm BH}$) at the center of Abell~262 is small
(table~\ref{tab:pro}). This may lead to a lower efficiency of power
generation, which may explain its peculiarity. Alternatively, the AGN
may have been at a ``pre-feedback'' phase \citep{2023arXiv230304821U}.

\subsection{The origin of the mm/submm continuum emission}

So far, the origin of the continuum emission from the AGNs in massive
elliptical galaxies has not been much studied in the mm/submm band. We
study the spectral index of the continuum emission $\beta_{\rm con}$ by
comparing the fluxes between the upper and lower sidebands of the CO
lines (table~\ref{tab:result}). For most galaxies, the frequency difference between the sidebands
is $\Delta\nu_{\rm ul}\sim 15$~GHz. However, it
is only $\sim 2$~GHz for 2A0335$+$096 and PKS~0745$-$191, and their
indices cannot be well constrained. Table~\ref{tab:result} shows that the indices are $\beta_{\rm con}< 0$ except for NGC~5044, which indicates
that the emission is likely to be dominated by synchrotron radiation and
the contribution of dust emission is minor. Thus, the assumption of
$\gamma=-0.75$ in section~\ref{sec:MLcon} seems to be reasonable.

\begin{figure}
 \begin{center}
  \includegraphics[width=8cm]{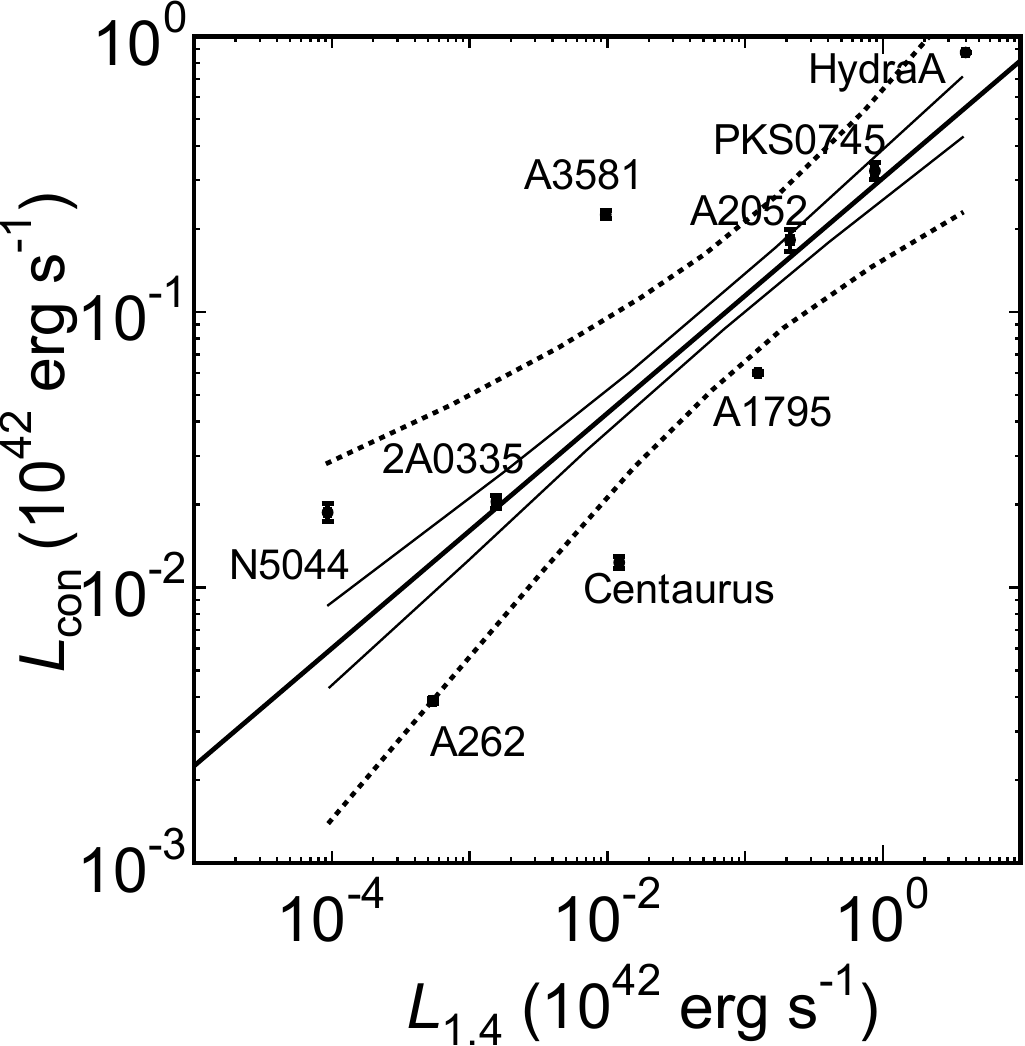} 
 \end{center}
\caption{The luminosity ($L_{\rm con}$) in the mm/submm band compared
with that at $1.4$~GHz ($L_{\rm 1.4}$). The thick solid line shows the
best-fitting power-law model that is determined using the BCES
estimator. The thin solid and dotted lines show the $1\:\sigma$ and
$3\:\sigma$ confidence ranges, respectively.}\label{fig:L14-Lcon}
\end{figure}

\begin{figure}
 \begin{center}
  \includegraphics[width=8cm]{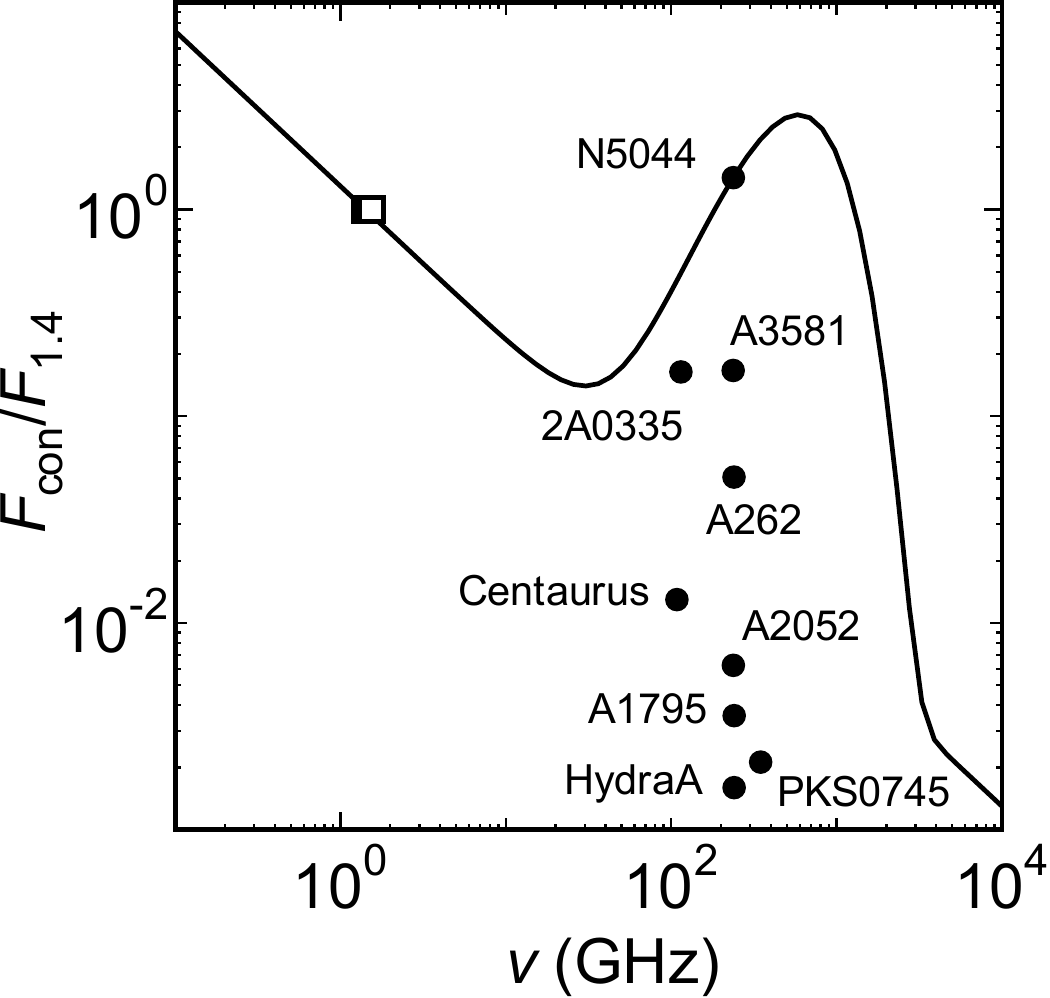} 
 \end{center}
\caption{Spectra of continuum emission. Filled circles show $((1+z)\nu_{\rm con}, F_{\rm con}/F_{1.4})$ and open squares show $((1+z)\nu_{1.4}, 1)$; the latter is almost independent of clusters. The solid line represents a power-law + modified blackbody spectrum reproducing the observations for NGC~5044 (see text). The errors of observation and the effects of $1+z$ are ignorable.}\label{fig:SED}
\end{figure}

Figure~\ref{fig:L14-Lcon} compares the continuum luminosities at
$\nu\sim \nu_{10}= 115.3$~GHz ($L_{\rm con}$) with those at $1.4$~GHz
($L_{1.4}$) estimated in section~\ref{sec:MLcon}. Those two luminosities
are positively correlated and the correlation coefficient is 0.823 with
$p$-value 0.0059.  The correlation among the 9 objects can be
represented by a power-law model of the form:
\begin{equation}
 \log\left(\frac{L_{\rm con}}{10^{42}\rm erg\: s^{-1}}\right) = A_3\log\left(\frac{L_{\rm 1.4}}{10^{42}\rm erg\: s^{-1}}\right) + B_3\:.
\end{equation}
The BCES $(Y|X)$ estimator gives $A_3=0.43\pm 0.10$ and $B_3=-0.51\pm
0.18$. We have confirmed that there is no correlation between fluxes
$F_{\rm con}$ and $F_{\rm 1.4}$ (the correlation coefficient is -0.2567
and $p$-value is 0.48). Thus, it is unlikely that the $L_{\rm
1.4}$--$L_{\rm con}$ relation with $A_3\neq 1$ is due to redshift
selection effects \citep{1983ApJ...269..400F}. The fact that $A_3$ is
significantly smaller than one means that the ratio $L_{\rm con}/L_{\rm
1.4}\propto L_{1.4}^{A_3-1}$ increases as $L_{\rm 1.4}$ decreases.  This
suggests that the mm/submm continuum emission from the AGNs with large
$L_{\rm 1.4}$ is pure synchrotron radiation, while that from the AGNs
with smaller $L_{\rm 1.4}$ is contaminated by additional dust
emission. 
Figure~\ref{fig:SED} shows the ratios $F_{\rm con}/F_{\rm 1.4}$
and illustrates a possible continuum spectrum (solid line). We assume
that the spectrum is simply represented by a combination of power-law (synchrotron) and modified
blackbody (dust) components:
\begin{equation}
 S(\nu) = N_{\rm AGN}\nu^{\gamma} + N_{\rm BB}(1-e^{-(\nu/\nu_0)^b})B_\nu(\nu,T)\:,
\end{equation}
where $\gamma=-0.75$, $\nu_0=1.5$~THz, $b=1.5$, and $B_\nu(\nu,T)$ is a
blackbody distribution (e.g. \cite{2019A&A...621A..27F}). The dust
temperature is assumed to be $T=10$~K, which is obtained for massive
elliptical galaxies at cluster centers
(e.g. \cite{2019ApJ...879..103F}). The relatively low temperature may be
because the surface area of the molecular gas is small for a given mass
and it is well shielded \citep{2019ApJ...879..103F}. If $T>10$~K, the
peak frequency of dust emission shifts to a higher frequency. The
constants $N_{\rm AGN}$ and $N_{\rm BB}$ are adjusted to reproduce the
observations of NGC~5044.  Figure~\ref{fig:SED} suggests that, except
for NGC~5044, the contribution of the dust component at $\sim
100$--300GHz is small, which is consistent with observations of local
AGNs \citep{2022ApJ...938...87K}.

\section{Conclusion}

We have studied the relation between the circumnuclear molecular gas and
the activities of the AGNs in the massive elliptical galaxies at the
center of galaxy clusters. We analyzed ALMA archival data for 9 galaxies
and obtained the circumnuclear molecular gas mass at $r<500$~pc from the
AGNs. We found that the mass has a correlation with the jet power of the
AGNs. Since the jet power is estimated from the size of X-ray cavities,
it should be the average power over the cavity formation period. Thus,
the correlation indicates that the circumnuclear gas is responsible for
long-term AGN activities ($\sim 10^7$~yr). We also estimated the mass
accretion rate of the cold gas assuming that the angular momentum
transfer is caused by turbulence. We showed that the accretion rate is
also correlated with the jet power. On the other hand, radio continuum
emissions from the AGNs at 1.4~GHz and $\sim 100$--300~GHz are not
correlated with the mass of circumnuclear molecular gas. This means that
the gas is not directly associated with the current AGN activity. The
spectra of the mm/submm ($\sim 100$--300~GHz) continuum emission
indicate that they are mostly synchrotron radiation. However, comparison
with 1.4~GHz luminosity suggests that dust emission is contaminated for
lower luminosity AGNs.



\begin{ack}
We thank the reviewer for helpful comments that improved the paper. We
also thank the EA ALMA Regional Center (EA-ARC) for their support. This
work was supported by NAOJ ALMA Scientific Research Grant Code 2022-21A,
and JSPS KAKENHI Grant Number JP20H00181, JP22H01268, JP22H00158 (Y.F.),
JP20K14531, JP21H04496 (T.I.), JP19K03918 (N.K.), JP21H01137, JP18K03709
(H.N.). This paper makes use of the following ALMA data:
ADS/JAO.2011.0.00735.S, ADS/JAO.2012.1.00837.S, ADS/JAO.2012.1.00837.S,
ADS/JAO.2015.1.00598.S, ADS/JAO.2015.1.00623.S, ADS/JAO.2015.1.00627.S,
ADS/JAO.2015.1.00644.S, ADS/JAO.2015.1.01198.S,
ADS/JAO.2016.1.01214.S. ALMA is a partnership of ESO (representing its
member states), NSF (USA) and NINS (Japan), together with NRC (Canada),
MOST and ASIAA (Taiwan), and KASI (Republic of Korea), in cooperation
with the Republic of Chile. The Joint ALMA Observatory is operated by
ESO, AUI/NRAO and NAOJ. Data analysis was in part carried out on the
Multi-wavelength Data Analysis System operated by the Astronomy Data
Center (ADC), National Astronomical Observatory of Japan.
\end{ack}

%
%


\end{document}